\documentclass[pra,twocolumn,floatfix,superscriptaddress]{revtex4-1} %reprint
\usepackage{etex}
\usepackage{amsmath}
\usepackage{bm}
\usepackage{bbm}
\usepackage{listings}
\usepackage{graphicx}
\usepackage{epsfig}
\usepackage{color}
\usepackage{multirow}
\usepackage[colorlinks]{hyperref}
\usepackage{fancyhdr}
\usepackage{calc}
\usepackage{natbib} %[numbers]
\usepackage{bibentry}

\newcommand{\dt}[1]{\frac{{\mathrm d} {#1}}{{\mathrm d}t}}
\def\br{\mathbf{r}}
\def\bra#1{\langle{#1}\rvert}%{\mathinner{\langle{#1}\rvert}}
\def\ket#1{\lvert{#1}\rangle}%{\mathinner{\lvert{#1}\rangle}}
\def\Braket#1#2{\mathinner{\langle{#1}\! \mid\! {#2} \rangle}}
\newcommand{\erf}[1]{Eq.~(\ref{#1})}
\newcommand{\frf}[1]{Fig.~\ref{#1}}
\newcommand{\srf}[1]{Sec.~\ref{#1}}
\newcommand{\nn}{\nonumber}
\newcommand{\mbf}[1]{\mathbf{#1}}
\newcommand{\op}[2]{\ket{#1}\bra{#2}}
\newcommand{\expt}[1]{\langle{#1}\rangle}
\newcommand{\dg}{^\dagger}
\newcommand{\smallfrac}[2]{\mbox{$\frac{#1}{#2}$}}
\newcommand{\Tr}{\mbox{Tr}}
\newcommand{\expect}[1]{\big\langle #1 \big\rangle}
\newcommand{\eff}{\text{eff}}

\newcommand{\half}{\smallfrac{1}{2}}

\newcommand{\oneD}{{\rm 1D}}
\newcommand{\vac}{{\rm vac}}
\newcommand{\cav}{{\rm cav}}
\newcommand{\inp}{{\rm in}}
\newcommand{\out}{{\rm out}}
\newcommand{\inter}{{\rm int}}
\newcommand{\scs}{{\rm SCS}}
\newcommand{\fwd}{+}
\newcommand{\bwd}{-}
\newcommand{\trans}{+}
\newcommand{\refl}{-}

\newcommand{\der}[1]{\frac{d {#1}}{dt}}
\newcommand{\unittens}{\tensor{\mathbf{I}}}
\newcommand{\poltens}{\hat{\tensor{\boldsymbol{\alpha}}}}
\newcommand{\varz}{\Delta J_3^2}
\newcommand{\jx}{\hat{J}_1}
\newcommand{\jz}{\hat{J}_3}
\newcommand{\shotnoise}{\Delta \mathcal{M}^2 |_{\rm SN}} 
 
\newcommand{\polcomp}{\hat{K}} % p,p' component of the tensor polarizability

\newcommand{\Eamp}{\mathcal{F}_0^{(+)}}
\newcommand{\charpol}{\alpha_0(\Delta_{f\!f'})}
\newcommand{\qaxis}{\mathbf{e}_{\tilde{z}}}
\newcommand{\qangle}{\varphi}
\newcommand{\magic}[1]{\tilde{\omega}_{#1}}
\newcommand{\chiN}{\chi_{N}}

\newcommand{\chieff}{\chi_{\raisebox{-.1pt}{\tiny $J_3$}}}

\newcommand{\gammauu}{\gamma_{\uparrow \rightarrow \uparrow}}
\newcommand{\gammadd}{\gamma_{\downarrow \rightarrow \downarrow}}
\newcommand{\gammaud}{\gamma_{\uparrow \rightarrow \downarrow}}
\newcommand{\gammadu}{\gamma_{\downarrow \rightarrow \uparrow}}
\newcommand{\gammau}{\gamma_{\uparrow}}
\newcommand{\gammad}{\gamma_{\downarrow}}

\newcommand{\Abir}{A_N}

\newcommand{\eigenf}{\mbf{f}_\eta}
\newcommand{\eigenfp}{\mbf{f}_{\eta'}}
\newcommand{\eigeng}{\mbf{g}_\eta}
\newcommand{\eigengp}{\mbf{g}_{\eta'}}

\newcommand{\awg}{\hat{a}_{b,p}(\omega)}
\newcommand{\awr}{\hat{a}_{m,p}(\omega,\beta)}

\begin{document}
\title{Dispersive response of atoms trapped near the surface of an optical nanofiber with applications to quantum nondemolition measurement and spin squeezing}
\author{Xiaodong Qi}
\affiliation{Center for Quantum Information and Control, University of New Mexico, Albuquerque, New Mexico 87131, USA}
\author{Ben Q. Baragiola}
\affiliation{Center for Quantum Information and Control, University of New Mexico, Albuquerque, New Mexico 87131, USA}
\author{Poul S. Jessen}
\affiliation{Center for Quantum Information and Control, University of Arizona, Tucson, Arizona 87521, USA}
\author{Ivan H. Deutsch}
\affiliation{Center for Quantum Information and Control, University of New Mexico, Albuquerque, New Mexico 87131, USA}
\date{\today}

%================================================================%
\begin{abstract}
We study the strong coupling between photons and atoms that can be achieved in an optical nanofiber geometry when the interaction is dispersive.  While the Purcell enhancement factor for spontaneous emission into the guided mode does not reach the strong-coupling regime for individual atoms, one can obtain high cooperativity for ensembles of a few thousand atoms due to the tight confinement of the guided modes and constructive interference over the entire chain of trapped atoms. We calculate the dyadic Green's function, which determines the scattering of light by atoms in the presence of the fiber, and thus the phase shift and polarization rotation induced on the guided light by the trapped atoms.  The Green's function is related to a full Heisenberg-Langevin treatment of the dispersive response of the quantized field to tensor polarizable atoms.  We apply our formalism to quantum nondemolition (QND) measurement of the atoms via polarimetry.  We study shot-noise-limited detection of atom number for atoms in a completely mixed spin state and the squeezing of projection noise for atoms in clock states.  Compared with squeezing of atomic ensembles in free space, we capitalize on unique features that arise in the nanofiber geometry including anisotropy of both the intensity and polarization of the guided modes.  We use a first principles stochastic master equation to model the squeezing as function of time in the presence of decoherence due to optical pumping.  We find a peak metrological squeezing of $\sim 5$ dB is achievable with current technology for $\sim 2500$ atoms trapped 180 nm from the surface of a nanofiber with radius $a=225$ nm.  
\end{abstract}

\maketitle

%===================INTRODUCTION=====================%
\section{Introduction}

Strong coupling between atoms and photons is at the heart of many quantum information processing protocols including efficient generation of remote entanglement \cite{duan_long-distance_2001, julsgaard_experimental_2001}, quantum data storage and retrieval \cite{ eisaman_electromagnetically_2005}, and QND measurements \cite{eckert_quantum_2008}.  
From a general perspective, strong coupling arises when atoms radiate predominantly into the electromagnetic field mode that defines the quantum atom-light interface.  
For an individual atom, the strong coupling regime is attained via the Purcell effect, whereby the boundary conditions of nearby dielectrics and/or conductors enhance radiation into a desired mode relative to all other modes.  
This can be achieved with Fabry-Perot cavities (cavity QED) \cite{miller_trapped_2005} and/or via nanophotonic structures engineered such that the radiation is predominantly into a specified mode \cite{manga_rao_single_2007, hakuta_manipulating_2012, hung_trapped_2013}.  The Purcell enhancement factors for emission into a guided or cavity mode scale respectively as  $ \Gamma_{\oneD}/\Gamma_{\vac} \sim \sigma_0/A$ and  $\Gamma_{\cav}/\Gamma_{\vac} \sim   Q \lambda^3/V \sim \mathcal{F}  \sigma_0/A$.  Here $\Gamma_{\vac}$ is the free space spontaneous emission rate, $\sigma_0 \propto \lambda^2$ is the resonant absorption cross section, $Q$, $V$, and $\mathcal{F}$,  are the cavity quality factor, volume, and finesse respectively, and $A$ is the effective area of the cavity or guided mode that couples to the atom.  The strongest coupling occurs on resonance, and thus much effort has been devoted to developing the largest possible $\Gamma_{\cav}$ and $\Gamma_{\oneD}$ through ultra-high-$Q$, small-volume resonators \cite{raimond_manipulating_2001, wallraff_strong_2004, miller_trapped_2005} and through nanophotonic plasmonic \cite{dzsotjan_quantum_2010, tame_quantum_2013}, metamaterial \cite{yao_ultrahigh_2009}, and dielectric \cite{hung_trapped_2013, goban_atomlight_2014} waveguides.   

In free space, where there is no Purcell enhancement, strong coupling can be achieved via the {\em cooperativity} of atomic ensembles.  
This is most naturally implemented in a dispersive regime, off resonance, where light elastically scattered from the ensemble constructively interferes to match the mode of an exciting paraxial probe \cite{baragiola_three-dimensional_2014}.  The cooperativity per atom in a typical paraxial beam is small, $\Gamma_{\oneD}/\Gamma_{\vac} \sim \sigma_0/A  \sim 10^{-6}$.  
The total cooperativity, however, can be significant for sufficiently large ensembles, e.g. $N_A \sim  10^7$ atoms. The key parameter that characterizes cooperativity is the total resonant optical density of the ensemble, OD $= N_A (\sigma_0/A)$.  
Such strong cooperativity in free space has been employed in a variety of applications including quantum memory for storage of photonic states \cite{chaneliere_storage_2005} and the generation of squeezed states of the collective spin of the ensemble via quantum nondemolition (QND) measurement \cite{kuzmich_generation_2000, appel_mesoscopic_2009, takano_spin_2009, sewell_magnetic_2012}.   

%========= FIGURE: Geometry (coupling strength, magic wavelength, area and detuning) =========%
\begin{figure}
\includegraphics[scale=0.80]{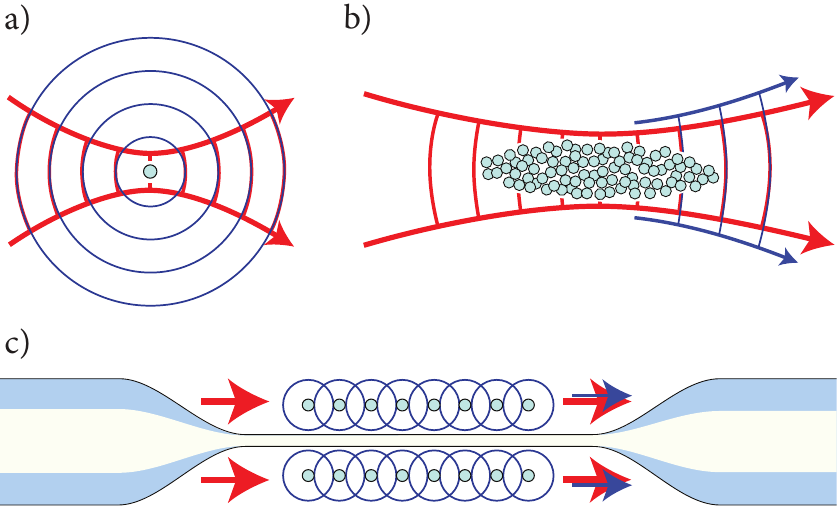}
\caption{Cooperativity and mode-matching for various atom-light geometries. (a) The  beam area at the waist of a tightly focused beam is closely matched with the atomic scattering cross section, but the scattered light of a single atom is poorly mode-matched with the probe. (b) A paraxial beam probing a rarefied atomic cloud whose scattered radiation interferes constructively in the forward direction. (c) Atoms trapped in a 1D optical lattices  near the surface of an optical nanofiber interacting with a fiber-guided probe. The tight confinement and automatic mode matching that accompanies scattering into the guided mode leads to strong cooperativity in the atom-photon interaction.}\label{Fig::ModeMatching}
\end{figure}
%=============================================

A particular system that combines the elements above consists of cold atoms trapped in the evanescent field of the guided mode of a tapered optical nanofiber with a subwavelength diameter \cite{vetsch_optical_2010, lacroute_state-insensitive_2012, balykin_quantum_2014, grover_photon-correlation_2015, Lee2015Inhomogeneous} (see \frf{Fig::ModeMatching}).  
The typical resonant OD per atom, or OD/$N_A$, in the nanofiber ($\sigma_0/A \sim  10^{-2}$) is boosted by orders of magnitude over free space for paraxial beams. 
However, one cannot reach the strong coupling regime where $\Gamma_{\oneD}$ is on the order of $\Gamma_{\vac}$ as is possible in engineered nanophotonic waveguides, such as those arising in photonic crystals \cite{hung_trapped_2013}, where atoms can be trapped at positions of peak intensity of the field.  
One can, however, achieve strong cooperativity in the dispersive regime with a moderately sized ensemble.  When compared to free space, all light scattered into the guided mode is automatically mode matched, and thus, given the relatively large ratio $\sigma_0/A$, one can achieve high OD with only a few thousand atoms (see \frf{Fig::ModeMatching}).  
Such strong cooperativity opens the door to new regimes to create non-Gaussian quantum states of the ensemble \cite{dubost_efficient_2012} and potentially to implement nonlinear optics at the level of a few photons \cite{spillane_observation_2008, pittman_ultralow-power_2013, oshea_fiber-optical_2013}.

One-dimensional optical lattices in nanofibers based on multiple co- and counter-propagating trapping beams have been loaded with up to several thousand alkali atoms \cite{vetsch_optical_2010, lacroute_state-insensitive_2012}.
This has proved a fruitful platform for quantum information processing.  
The anisotropic nature of the strong atom-light coupling has been exploited for control of internal atomic states \cite{mitsch_exploiting_2014}, enhanced coupling into a preferred propagation direction \cite{petersen_chiral_2014, mitsch_quantum_2014}, and optical switching \cite{oshea_fiber-optical_2013}. 
Off resonance, dispersive coupling has allowed for non-destructive atom counting \cite{dawkins_dispersive_2011, beguin_generation_2014} 
and storage of fiber-guided light \cite{gouraud_demonstration_2015, sayrin_storage_2015}.
Recent demonstrations of photonic crystal cavities fabricated on the nanofiber \cite{wuttke_nanofiber_2012, nayak_optical_2014, schell_highly_2015} promise further enhanced atom-light coupling.

In this paper we study the quantum atom-light interface in the dispersive regime for an optical nanofiber geometry.  We focus here on the coupling between the atomic spin and light polarization induced by the elastic scattering of photons by tensor-polarizable cesium atoms trapped near the surface of the nanofiber.  This provides an entangling interaction that can be employed to generate spin squeezing via QND measurement.  Our analysis unifies a variety of different approaches found in the literature, including direct calculation of the dyadic Green's function for photon scattering \cite{sakoda_optical_1996, dung_spontaneous_2000, sondergaard_general_2001, klimov_spontaneous_2004, wubs_multiple-scattering_2004, fussell_decay_2005, manga_rao_single_2007, dzsotjan_quantum_2010} and the input-output formalism studied for one-dimensional field theories based on Heisenberg-Langevin equations \cite{gardiner_input_1985, blow_continuum_1990, shen_coherent_2005, le_kien_spontaneous_2005, le_kien_correlations_2008, fan_input-output_2010}.

The remainder of this article is organized as follows.  
In \srf{Sec::GreensFunction} we solve for the mode decomposition of the dyadic Green's function which determines the electric field scattered by a point dipole near the surface of the nanofiber.  
This allows us to calculate the phase shift and polarization transformation for fiber-guided photons induced by tensor-polarizable atoms in the dispersive regime.  
We connect this with a fully quantum mechanical treatment based on a Heisenberg-Langevin picture in \srf{Sec::HeisenbergLangevin}.  
The formalism we develop is used in \srf{Sec::QNDMeasurement} to study QND measurement of atoms based on polarization spectroscopy. 
We consider shot-noise-limited atom detection as well as measurement-backaction-induced squeezing of spin projection noise.  
We study squeezing of the collective pseudospin associated with ensembles of atoms in the atomic clock state and calculate its dynamics based on a first principles stochastic master equation that includes both the effects of QND measurement as well as decoherence due to optical pumping.  
We conclude with a summary and outlook for future research in \srf{Sec::Conclusion}.

%========GREEN'S FUNCTION AND INPUT/OUTPUT RESPONSE=========%
\section{Dyadic Green's function and input-output field response} \label{Sec::GreensFunction}

Given a point particle with tensor polarizability $\tensor{\boldsymbol{\alpha}}$ at position $\br'$ near the surface of a nanofiber, the field  at frequency $\omega_0$ is given by the solution to the wave equation, 
	\begin{align}\label{Eq::WaveEquationSource}
		\left[ - \!\nabla\!\!\times\!\nabla\!\!\times \!+  n^2\!(\br)k_0^2 \right]\! \mathbf{E}(\br) &\!=\! -4\pi  k_0^2 \delta^{(3)}\!(\br\!-\!\br')\,  \tensor{\boldsymbol{\alpha}}\!\cdot\! \mathbf{E}(\br),
	\end{align}
where $k_0=\omega_0/c$ and $n(\mbf{r})$ is the spatially varying index of refraction that describes the fiber; Gaussian-cgs units are used throughout.  
For an asymptotic input field $\mathbf{E}_{\inp}(\br)$, the scattering solution to \erf{Eq::WaveEquationSource} is given by the Lippmann-Schwinger equation \cite{wubs_multiple-scattering_2004},
\begin{subequations}
	\begin{align}
		\mathbf{E}_{\out}(\br) &=\mathbf{E}_{\inp}(\br)+\tensor{\mathbf{G}}^{(+)}(\br , \br'; \omega_0)\cdot 
\tensor{\boldsymbol{\alpha}}\cdot \mathbf{E}_{\out}(\br')\\
		&\approx \mathbf{E}_{\inp}(\br)+ \tensor{\mathbf{G}}^{(+)}(\br , \br'; \omega_0) \cdot 
\tensor{\boldsymbol{\alpha}}\cdot \mathbf{E}_{\inp}(\br'), \label{Eq::ScatteredField}
	\end{align}
\end{subequations}
where in \erf{Eq::ScatteredField} we have made the first Born approximation valid for weak scattering. The fundamental object that fully characterizes the scattered radiation as well as the energy level shift and modified decay rate of a scatterer near the dielectric is the dyadic Green's function, $\tensor{\mathbf{G}}(\br, \br';\omega_0)$. This determines the scattered field from a point dipole at $\br'$, $\mathbf{E}_{\rm scat}(\br)= \tensor{\mathbf{G}}^{(+)}(\br , \br'; \omega_0)\cdot \mathbf{d}$, and satisfies the equation of motion,
	\begin{align} \label{Eq::GreensDiffEq}
		\!\!\!\!\!\!\left[ -\!\nabla\!\times\!\nabla\!\times + n^2(\mbf{r}) k_0^2 \right]\! \tensor{\mathbf{G}}(\br,\! \br';\omega_0) &\!=\! -4\pi 
k_0^2 \delta^{(3)}(\mathbf{r}\!-\!\mathbf{r}')\! \unittens,
	\end{align}
where $\unittens$ is the unit tensor.   

The solution for the Green's function $\tensor{\mathbf{G}}(\br ,\br' ; \omega_0)$, following from Maxwell's equations,  
has been studied previously \cite{sakoda_optical_1996,sondergaard_general_2001,wubs_multiple-scattering_2004}.  As we are interested here in the forward-scattered components that lead to phase shifts and polarization transformations, we directly calculate $\tensor{\mathbf{G}}(\br ,\br' ; \omega_0)$ through a decomposition into normal modes.  A complete set of eigenmodes in the presence of lossless, spatially inhomogeneous dielectric are defined according to the procedure of Glauber and Lewenstein~\cite{glauber_quantum_1991}.  We seek the eigenmodes $\eigenf(\mathbf{r})$, indexed by $\eta$, that satisfy the homogeneous wave equation in the absence of sources, i.e., \erf{Eq::WaveEquationSource} for $\tensor{\boldsymbol{\alpha}} = 0$ with the eigen-wavenumber $k_0 \rightarrow k_\eta$.  To do so, one defines functions $\eigeng(\mbf{r}) \equiv n(\br) \eigenf(\mbf{r})$ that form a complete basis, as they are eigenfunctions of the Hermitian operator, $\mathcal{H}(k_0) = -\frac{1}{n(\br)} \nabla\times\nabla\times \frac{1}{n(\br)} + k_0^2$, according to $\mathcal{H}(k_0)  \eigeng(\mbf{r}) = \lambda_\eta \eigeng(\mbf{r})$. The eigenvalue, $\lambda_\eta= (\omega_0^2-\omega_\eta^2)/c^2$, determines the wavenumber for a given mode at frequency $\omega_\eta$.  We are interested specifically in the generalized transverse functions satisfying $\nabla\cdot [ n(\mathbf{r}) \eigeng(\br) ] = 0$ with eigenvalues $\lambda_n \neq 0$ \cite{wubs_multiple-scattering_2004}. These fall into two categories, guided ($\eta = \mu$) and unguided ($\eta = \nu$) modes, which together form a complete, orthonormal set for transverse vector functions,
	\begin{align}
	&\int\!\! \mathrm{d}^3\br \, \eigeng^*(\mbf{r}) \!\cdot\! \eigengp(\mbf{r})  =\!\! \int\!\! \mathrm{d}^3\br \, n^2(\br) \eigenf^* (\br)\!\cdot\!  \eigenfp(\br) =\delta_{\eta, \eta'},\label{Eq::Orthogonality}\\
	&\!\!\!\!\!\!\!\sum_\eta \!\mathbf{g}_\eta\!(\!\br\!) \mathbf{g}_\eta^*\!(\!\br'\!) \!=\!\!\!  \sum_\mu\! \mathbf{g}_\mu\!(\!\br\!) \mathbf{g}_\mu^*\!(\!\br'\!)  \!+\!\! \sum_\nu\! \mathbf{g}_\nu\!(\!\br\!) \mathbf{g}_\nu^*\!(\!\br'\!)  \!=\!\tilde{\delta}^{(\!T\!)}\!(\!\br\!-\!\br'\!)\!  \unittens, \label{Eq::Completeness}
	\end{align}
where $\tilde{\delta}^{(T)}(\br-\br')$ is the delta function for generalized transverse vector fields \footnote{The functions $\eigeng(\br)$ are not strictly transverse because of the spatial variation of the index of refraction, $n(r_\perp)$.  These modes do, nonetheless, constitute the far-field radiated by the dipole. For further details see Refs. \cite{sakoda_optical_1996, wubs_multiple-scattering_2004} }.  It follows that the generalized transverse dyadic Green's function can be decomposed in terms of the eigenfunctions~\cite{sakoda_optical_1996, sondergaard_general_2001}
	\begin{align}
		\tensor{\mathbf{G}}^{(T)}(\br,\br'; \omega_0) &= -4\pi \sum_{\eta} \frac{  \omega_0^2 \eigenf (\br) 
\eigenfp^* (\br')}{\omega_0^2-\omega_\eta^2},
	\end{align}
where the eigenvalues appear as $\omega_\eta^2 = c^2 k_\eta^2$.  The sum includes both guided and unguided contributions. We focus here on the guided-mode contribution to the Green's function. 

We treat an optical nanofiber of radius $a$ with step-index profile,
	\begin{align} \label{Eq::IndexofRefraction}
		n(r_\perp) = \Big\{  
			\begin{array}{l l} n_1 & \quad r \leq a \\
						 n_2 & \quad r > a 
		\end{array},
	\end{align}
for a silica core ($n_1 = 1.4469$)~\cite{kien_field_2004} and infinite vacuum cladding ($n_2 = 1$).  For a cylindrically symmetric dielectric the guided modes are $\mathbf{f}_\mu (\br) = \mathbf{u}_\mu (\br_\perp) e^{i\beta z}/\sqrt{2 \pi}$, with indices $\mu=\{j, \beta , p\}$ for the $j^{th}$ guided mode with propagation constant $\beta$ at frequency $\omega_\mu=\omega(\beta)$ and polarization $p$.  The transverse mode functions are normalized according to $\int d^2 \mbf{r}_\perp \, n^2(r_\perp)\mathbf{u}^*_\mu (\br_\perp) \cdot \mathbf{u}_{\mu'} (\br_\perp)\big|_{\beta = \beta'} = \delta_{j,j'}\delta_{p,p'}$ and have units $1/\sqrt{A}$ \cite{le_kien_anisotropy_2014}.  Two convenient guided-mode bases are the quasilinear and quasicircular polarization modes, described in Appendix \ref{Appendix::ModeFunctions} \cite{kien_field_2004}.  

We consider nanofibers that support only the lowest HE$_{11}$ guided modes at the relevant frequency $\omega_0$ \cite{snyder_optical_1983}, and thus we drop the mode index $j$.  In this case there are four guided modes: two polarizations $p$, each with propagation constant $\beta(\omega_0) = \pm\beta_0$ corresponding to forward and backward propagation.  The guided-mode contribution to the dyadic Greens function is then
	\begin{align} \label{Eq::GreensEigenmodes}
		\tensor{\mathbf{G}}\!_g\!(\!\br\!,\br'\!; \omega_0\!) \!=&\!\! \int_{\!-\infty}^\infty \!\!\!\!\!\! \mathrm{d} \beta \!\sum_{p} \!\!
\frac{-2\omega_0^2}{\omega_0^2\!\!-\!\omega^2(\!\beta\!)} \mathbf{u}_{\beta\!,p} (\!\br\!_\perp\!)\mathbf{u}^*_{\beta\!,p} 
(\!\br_{\!\perp}^\prime\!) e^{i\beta(\!z\!-\!z'\!)}\!,
	\end{align}
where $ \omega(\beta)$ is the frequency of the guided HE$_{11}$ for a given $\beta$.  

For $z>z'$ ($z<z'$), the contribution of the guided modes to the retarded (causal) Green's function is found by the 
usual displacement of the pole on the positive (negative) $\beta$-axis into the upper (lower) half of 
the complex plane. The result for $z \neq z'$ is \cite{manga_rao_single_2007}
	\begin{align} 
		&\tensor{\mathbf{G}}^{(+)}_g(\br,\br'; \omega_0)\nn\\ 
		\!=& 2\pi i\!\! \sum_{b,p}\!\!  {\rm Res}\vert_{\beta\! =b\beta_0} \!\!
\left[\!\frac{-2 \omega_0^2 }{ \omega_0^2\!\!-\!\!\omega^2(\beta)}\!\right]\!   \mathbf{u}_{b\beta_0\!, p} 
(\!\br\!_\perp\!)\mathbf{u}^*_{b\beta_0\!, p} (\!\br_{\!\perp}^\prime\!)e^{ib \beta_0 \!(\!z\!-\!z'\!)} \nonumber \\
\!= & 2\pi i \frac{\omega_0}{v_g }\! \sum_{b,p}\! \mathbf{u}_{b, p} (\br\!_\perp)\mathbf{u}^*_{b, p} 
(\br_{\!\perp}^\prime) e^{i b\beta_0(\!z\!-\!z'\!)} \Theta \big( (\!z\!-\!z'\!)b \big), \label{Eq::GreensGuided}
	\end{align}
where $b=\pm$ indicates the propagation direction, $v_g= \vert d\omega/d\beta \vert_{\beta=\beta_0}$ is the group velocity at $\omega_0$, and $\Theta \big( b(z-z') \big)$ is a Heaviside function enforcing causality for the forward- and backward-scattered fields. In the second line, we have suppressed the label $\beta_0$ as it is implicit in the definition of the guided modes at frequency $\omega_0$. 

Radiative properties of a scatterer (the decay rate and energy level shift) are determined by evaluation of the dyadic Green's function at the source point $\mbf{r} = \mbf{r}'$ \cite{fussell_decay_2005}.  However, for $z=z'$ we cannot close the contour. Instead, we expand the resonant denominator in \erf{Eq::GreensEigenmodes} with the poles moved to yield the retarded (causal) response,
\begin{equation}
\frac{1}{(\omega_0\!\!+\! i\epsilon)^2\!-\!\omega^2(\beta)} \!=\!\frac{1}{2 \omega(\beta)}\!\left[\! \frac{1}{\omega_0\!\!+\! i 
\epsilon \!-\! \omega(\beta)} \!-\! \frac{1}{\omega_0\!\!+\! i \epsilon \!+\! \omega(\beta)} \right]\!,\nn
\end{equation}
 and employ the usual distribution identities \cite{sondergaard_general_2001},
\begin{equation}
\!\!\!\!\lim_{\epsilon \!\rightarrow 0_+} \frac{1}{\omega_0 \!+\! i \epsilon \!\mp\! 
\omega(\beta)}=\mathcal{P}\left[\frac{1}{\omega \!\mp\! \omega(\beta)} \right] \!+\! i \pi \delta (\omega_0 \!\mp\! 
\omega(\beta)).
\end{equation}
Only the positive-frequency component contributes to the $\delta$-function, and it follows that the imaginary part of the Green's function at $\br = \br'$ that determines the resonant Purcell enhancement of spontaneous emission into the guided modes is \cite{dung_spontaneous_2000, fussell_decay_2005, chen_finite-element_2010}
	\begin{equation}\label{Eq::ImGreenLocal}
		{\rm Im} \big[\tensor{\mathbf{G}}^{(+)}_g\!(\br'\!\!,\br'; \omega_0\!\!=\!\omega\!_{eg}) \big] \!=\! \pi \frac{\omega\!_{eg}}{v_g }\!\! \sum_{b, p}\! 
		\mathbf{u}_{b\!, p} (\br_{\!\perp}^\prime)\mathbf{u}^*_{b\! , p} (\br_{\!\perp}^\prime),
	\end{equation}
where $\omega_{eg}$ is resonance frequency of the atomic scatterer.  The energy level shift of the scatterer due to its proximity to the dielectric is found from the real part of the Green's function at $\br = \br'$. 
To find the total modified spontaneous emission rate and energy level shift one must include the unguided radiation modes \cite{le_kien_spontaneous_2005} or employ other representations of the Green's function \cite{klimov_spontaneous_2004}.  

Equation (\ref{Eq::GreensGuided}) is the central result from which we can calculate the dispersive response.  Consider a forward-propagating input field in the guided modes with frequency $\omega_0$, positive-frequency amplitude $\Eamp$, and arbitrary polarization, $\mathbf{E}^{(+)}_{\inp}(\br) = \Eamp  \mathbf{u}_{\rm in}(\br_\perp) e^{i \beta_0 z}$ dispersively coupled to an atom at position $\mathbf{r}'$.  The effective mode area at the atom's position is determined from the total cycle-averaged power transported along the nanofiber, $P_{{\rm in},z} = (v_g/2\pi) \int d^2\br \, n^2(r_\perp) |\mathbf{E}^{(+)}_{\inp}(\br) |^2$, and the intensity at the atom, $I_{\rm in}(\mathbf{r}') = (c/2\pi) |\mathbf{E}^{(+)}_{\inp}(\br') |^2$, via the relation \cite{domokos_quantum_2002},
 	\begin{align} \label{Eq::AreaIn}
 		A_{\rm in} \equiv \frac{P_{{\rm in}}}{I_{\rm in}(\mathbf{r}')} = \frac{1}{n_g |\mathbf{u}_{\rm in}(\mathbf{r}'_\perp)|^{2}},
	\end{align}
where $n_g\equiv c/v_g$ is the group index of refraction.    

Substitution of the guided-mode Green's function, \erf{Eq::GreensGuided}, into the Lippman-Schwinger equation, \erf{Eq::ScatteredField}, yields the transmitted (forward-scattered) and reflected (backward-scattered) output fields, $\mathbf{E}_{\out}(\br) = \Eamp \big[ \mathbf{u}_{\trans, \out} (\br_\perp) e^{i \beta_0 z} + \mathbf{u}_{\refl,\out} (\br_\perp) e^{-i \beta_0 z} \big]$, 
\begin{subequations}
	\begin{align}
		\mathbf{u}_{\trans, \out} (\br_\perp) &=  \sum_{p,p'}  \, c_{p} t_{pp'} \mathbf{u}_{\fwd, p'}(\br_\perp) \\ 
		\mathbf{u}_{\refl,\out} (\br_\perp) &=  \sum_{p,p'}  \, c_{p} r_{pp'} \mathbf{u}_{\bwd, p'}(\br_\perp),
	\end{align}
\end{subequations}
where we have decomposed the input into the polarization eigenmodes, $\mbf{u}_{\rm in}(\mbf{r}_\perp) = \sum_{p} c_{p} \mathbf{u}_{\fwd,p}(\br_\perp)$.  
For $z>z'$, the transmission and reflection matrices are 
\begin{subequations}
	\begin{align} \label{Eq::PolarizationTransformation}
		t_{pp'} =& \delta_{p,p'} +  2\pi i k_0 n_g \, \mathbf{u}^*_{+, p}(\br'_\perp) \cdot 
\tensor{\boldsymbol{\alpha}} \cdot \mathbf{u}_{+, p'}(\br'_\perp) , \\
		r_{pp'} =&  2\pi i k_0 n_g \, \mathbf{u}^*_{\bwd, p}(\br'_\perp) \cdot 
\tensor{\boldsymbol{\alpha}} \cdot \mathbf{u}_{\fwd, p'}(\br'_\perp) e^{2 i\beta_0 z'} , 
	\end{align} 
\end{subequations}
We focus here on the transmitted fields whose interference with the input field for $z>z'$ results in a phase shift and a polarization transformation.  
For weak scattering the diagonal terms, $t_{p p} \approx \sqrt{1-R_p}e^{i \delta \phi_p}$, determine the phase shift and attenuation induced on each polarization mode,
\begin{subequations}
	\begin{align}
		 \delta \phi_p &= \frac{2 \pi k_0}{A_{\rm in}} {\rm Re}(\alpha_{pp}),  \label{Eq::PhaseShift} \\
		R_p &=  \frac{4 \pi k_0}{A_{\rm in}} {\rm Im}(\alpha_{pp}) .\label{Eq::Attenuation} 
	\end{align} 
\end{subequations}
Here, the $\{p,p'\}$-element of the tensor polarizability is given by, $\alpha_{pp'} \equiv \mathbf{e}^*_{p'} \cdot \tensor{\boldsymbol{\alpha}}\cdot \mathbf{e}_{p}$, with unit vectors for each of the forward-propagating mode functions, $\mathbf{e}_{p}\equiv \mathbf{u}_{+,p}(\br'_\perp)/|\mathbf{u}_{+,p}(\br'_\perp)|$. 

The phase shift per atom, \erf{Eq::PhaseShift}, is modified over free space in two ways, both of which are captured by the effective mode area $A_{\rm in}$. First, although material dispersion in an optical fiber is negligible over the distances we consider, additional waveguide dispersion can lead to a significant reduction in the group velocity~\cite{hung_trapped_2013,goban_atomlight_2014}.  Such ``slow light" enhances the atom-photon coupling strength. 
In the nanofiber geometry this effect is moderate -- we calculated the group index to be $n_g \approx 1.40$. 
Second and more importantly, the tight spatial confinement as measured by OD/$N_A$ significantly increases the coupling strength over free space for every atom along the nanofiber, which yields strong cooperativity.
In contrast, in free space diffraction restricts the collective phase shift for an ensemble of atoms~\cite{tanji-suzuki_chapter_2011, baragiola_three-dimensional_2014}.  
For a Gaussian beam with beam waist $w_0$, the total phase shift induced by a collection of polarizable atoms will be $\delta \phi = N_{\eff} 2 \pi k_0 {\rm Re}({\alpha})/A$, where $A = \pi w^2_0/2$ is the beam area at the focus and $N_{\eff}$ is the effective number of atoms that radiate into this mode.  
One can couple strongly to few atoms at the center by tightly focusing the beam or couple weakly to many atoms by choosing a larger focal volume, but hence, smaller cooperativity per atom.  

The off-diagonal terms in the transmission matrix, \erf{Eq::PolarizationTransformation}, describe the polarization transformation. For example, if we take the polarization of the modes to be the quasilinear, $p = \{H,V\}$ as defined in \erf{Eq::QuasilinearModes}, then $t_{HV} \equiv \chi_{\rm Far}$ is the rotation angle of the Stokes vector on the Poincar\'{e} sphere corresponding to the Faraday effect \cite{hammerer_quantum_2010, deutsch_quantum_2010}.  
The phase difference in that basis, $\delta  \phi_H - \delta \phi_V$, corresponds to birefringence induced on the guided mode and $t_{HV}$ to Faraday rotation.  
Analyzed in the quasicircular polarization modes ($p=\pm$), given in \erf{Eq::QuasicircularModes}, the differential phase $\delta \phi_+ -\delta  \phi_-$ corresponds to Faraday rotation and $t_{+-}$ to birefringence.  
We make use of such polarization transformations as a means to nondestructively measure the atoms and generate collective spin squeezing.

%===================Heisenberg-Langevin Equations=====================%
\section{Heisenberg-Langevin-picture solution and atomic response} \label{Sec::HeisenbergLangevin}
	
The Lippmann-Schwinger solution, \erf{Eq::ScatteredField}, determines the input-output relation for linear atomic response given by the polarizability tensor $\tensor{\alpha}$.  
In this section we connect this with the fully quantum mechanical description of dispersive atomic response and input-output relations for the quantized guided modes.  
Following Ref. \cite{le_kien_spontaneous_2005}, we use a Heisenberg-Langevin approach for one-dimensional systems.  

The positive frequency component of the quantized electric field operator decomposes into guided and radiation (unguided) modes, $\hat{\mathbf{E}}^{(+)}=\hat{\mathbf{E}}_g^{(+)}+\hat{\mathbf{E}}_{r}^{(+)}$, where
\begin{subequations}
	\begin{align}
		\!\!\!\!\hat{\mathbf{E}}_g^{(\!+\!)}(\br) &= \sum_{b,p} \int_0^\infty\!\!\!\! \mathrm{d}\omega  \sqrt{\frac{ \hbar \omega}{ v_g}} \; \awg \mathbf{u}_\mu (\br\!_\perp) e^{i b\beta(\omega) z } ,\label{Eq::QuantizedElectricField} \\
		\!\!\!\!\hat{\mathbf{E}}_r^{(\!+\!)}(\br) &\!\!=\!\!\! \sum_{m\!,p}\!\! \int_0^{\infty}\!\!\!\!\!\! \mathrm{d}\omega\!\!\!   \int_{\!-kn_2}^{kn_2}\!\!\!\!\!\!\mathrm{d}\beta \, \sqrt{ \hbar \omega}\,\awr \mathbf{u}_\nu (\br\!_\perp) e^{i\beta(\!\omega\!) z }\!.
	\end{align}
\end{subequations}
The HE$_{11}$ guided modes are specified by $\mu =(\omega, b, p)$, where $\omega$ is the mode frequency,  $p$ is the polarization, and the propagation direction $b=\pm$ corresponds to wavenumber $b \beta (\omega)$.  The radiation modes are specified by  $\nu=(\omega, \beta, m, p)$, where $m$ is the azimuthal (angular momentum) quantum number, $p$ labels the two orthogonal polarizations, and longitudinal propagation constant $\beta$ can vary continuously from $-kn_2$ to $kn_2$, with $k = \omega/c$ \cite{sondergaard_general_2001,le_kien_spontaneous_2005}.  
The creation/annihilation operators satisfy the usual continuous-mode commutation relations, $[\hat{a}_\mu, \hat{a}^\dag_{\mu'} ] = \delta_{b,b'} \delta_{p,p'} \delta ( \omega - \omega ') $ and $[\hat{a}_\nu ,\hat{a}^\dag_{\nu'} ] = \delta_{m,m'} \delta_{p,p'} \delta ( \omega - \omega ')  \delta ( \beta - \beta') $.

The Hamiltonian for the system is
\begin{equation}
\hat{H} = \hat{H}_F+\hat{H}_A + \hat{H}_{\inter},
\end{equation}
where the free-field Hamiltonian decomposes into guided and unguided modes, 
	\begin{equation}
		\hat{H}_F = \!\sum_{b,p}\!\! \int_0^{\infty}\!\!\! \mathrm{d}\omega \, \hbar \omega \hat{a}^\dagger_\mu \hat{a}_\mu 
\!+\!\sum_{m,p}\! \int_0^{\infty}\!\!\! \mathrm{d}\omega\!\!\!  \int_{\!-k n_2}^{k n_2}\!\!\! \mathrm{d}\beta \, \hbar \omega 
\hat{a}^\dagger_\nu \hat{a}_\nu.
	\end{equation}
We consider here alkali atoms with ground and excited levels, $\{ \ket{g}=\ket{nS_{1/2}, f, m_f}\}$, $\{ \ket{e} =\ket{nP_{j'}, f', m_{f'}}\}$, where $\ket{f, m_f}$ denotes the hyperfine sublevels.  The free atomic Hamiltonian is
	\begin{equation}
		\hat{H}_A  = \sum_g E_g \hat{\sigma}_{gg} + \sum_e E_e \hat{\sigma}_{ee},
	\end{equation}
where $\hat{\sigma}_{ij} \equiv \ket{i}\bra{j}$.  In the rotating wave approximation, the atom-field interaction Hamiltonian is
	\begin{align}
		\hat{H}_{\inter} &\!=\! -\hat{\mathbf{d}}\!\cdot\! \hat{\mathbf{E}} =- \!\sum_{e,g}\! \left[ \hat{\mathbf{d}}_{eg}\!\!\cdot\! 
\hat{\mathbf{E}}^{(+)}(\br') \!+\! \hat{\mathbf{d}}_{ge}\!\!\cdot\! \hat{\mathbf{E}}^{(-)}(\br') \right],
	\end{align}
where the atomic dipole operator is projected between excited and ground subspaces, $\hat{\mathbf{d}}_{eg}= \hat{P}_e \hat{\mathbf{d}} \hat{P}_g $. The interaction Hamiltonian then takes the form, 
\begin{align}
	\hat{H}_{\inter} &= -\sum_{e,g} \left(\sum_{b,p} \int_0^{\infty}\!\!\!\!\mathrm{d}\omega \; \hbar g_{\mu, e,g}\, \hat{a}_\mu  \, 
		\hat{\sigma}_{eg}\right.\nonumber\\
	&\quad \left. + \sum_{m,p} \int_0^{\infty}\!\!\!\!\mathrm{d}\omega \! \int_{-kn_2}^{kn_2}\!\!\!\mathrm{d}\beta \,  \hbar 
g_{\nu, e,g}\, \hat{a}_\nu \, \hat{\sigma}_{eg}\right) + {H.c.},
	\end{align}
where the coupling constants for guided/radiation modes are
\begin{subequations} \label{Eq::CouplingConstants}
	\begin{align}
		\hbar g_{\mu, e,g} &= \sqrt{\frac{\hbar \omega}{ v_g  }}\, \bra{e} \hat{\mathbf{d}} \ket{g} 
\cdot\mathbf{u}_\mu ( \br'_\perp ) e^{i b \beta(\omega)z} , \\
		\hbar g_{\nu, e,g} &= \sqrt{  \hbar \omega } \, \bra{e} \hat{\mathbf{d}} \ket{g} \cdot \mathbf{u}_\nu ( \br'_\perp) e^{i\beta(\omega)z}  .
	\end{align}
\end{subequations}
The Heisenberg equations of motion are
\begin{subequations}
	\begin{align}
		\der{\hat{a}_\mu} &= -i\omega \hat{a}_\mu +i\sum_{e,g} g_{\mu, e,g}^* \hat{\sigma}_{ge} \label{eq:da},\\
		\der{\hat{a}_\nu} &= -i\omega \hat{a}_\nu +i\sum_{e,g} g_{\nu, e,g}^*  \hat{\sigma}_{ge}\label{eq:danu},\\
		\der{\hat{\sigma}_{ge}} &=\! -i\omega\!_{eg} \hat{\sigma}\!_{ge} 
			\!+\! i\!\!\int_0^{\infty}\!\!\!\!\!\! \mathrm{d}\omega\!\! \sum_{e'\!\!,g'}\!\! \bigg[\! \big(\delta\!_{ee'} \hat{\sigma}\!_{gg'} \!-\! \delta\!_{gg'} \hat{\sigma}\!_{e'\!e} \big) \label{Eq::dsigma}  \nn\\
		&\!\!\!\!\! \bigg\{ \sum_{b,p}  g_{\mu, e',g'}\hat{a}_\mu \!+\! \sum_{m,p} \!\int_{-kn_2}^{kn_2}\!\!\!\!\!\! \mathrm{d}\beta \; g_{\nu, e',g'} \hat{a}_\nu \bigg\} \bigg]. 
	\end{align}
\end{subequations}
Integrating the field equations, 
\begin{subequations}\label{eq:aout1}
\begin{align}
\hat{a}_\mu(t) &= \hat{a}_\mu(t_0) e^{-i\omega (t-t_0)} \nonumber\\
&\quad +i \sum_{e,g} g_{\mu,e,g}^* \int_{t_0}^t 
\mathrm{d} t' e^{-i\omega (t-t')}\hat{\sigma}_{ge}(t'), \label{Eq::aguidedEOM}
\end{align}
\begin{align}
\hat{a}_\nu (t) &= \hat{a}_\nu (t_0) e^{-i\omega (t-t_0)} \nonumber\\
&\quad +i \sum_{e,g} g_{\nu,e,g}^* \int_{t_0}^t \mathrm{d} 
t' e^{-i\omega (t-t')}\hat{\sigma}_{ge}(t'),
\end{align}
\end{subequations}
substituting into \erf{Eq::dsigma}, and making the usual Markov approximation \cite{le_kien_spontaneous_2005} gives an expression for the ground-excited coherences.   This yields
\begin{align}
&\dt{\hat{\sigma}_{ge}} \!=\!-i\omega\!_{eg} 
\hat{\sigma}\!_{ge}\!\!-\!\!\sum_{e'}\!\!\frac{\Gamma_{ee'}}{2}\hat{\sigma}\!_{ge'}  
\!+\!i\! \sum_{e'\!\!,g'}\!\!\bigg[\! (\delta\!_{e,e'} \hat{\sigma}\!_{gg'} \!-\! \delta_{g,g'} 
\hat{\sigma}\!_{e'\!e})\nn\\
&\int_0^{\infty}\!\!\!\!\!\mathrm{d}\omega \bigg\{\!\!\sum_{b,p} \! g_{\mu, e'\!\!,g'} \hat{a}_\mu (t_0) 
\!+\!\!\sum_{m,p}\!\!  \int_{\!-kn_2}^{kn_2}\!\!\!\!\!\!\!\!\!\mathrm{d}\beta  g_{\nu, e'\!\!,g'} \hat{a}_\nu(t_0) \bigg\} e^{\!-i\omega 
(t\!-\!t_0)} \!\bigg], \nonumber
\end{align}
where the decay rates of excited-populations and coherences are given by 
	\begin{align}
		\Gamma_{ee'} &= 2\pi \sum_{\mu,g} g_{\mu,e,g}g^*_{\mu,e',g} \vert_{\omega=\omega_{eg}}\nn\\
		&\quad +2\pi 
\sum_{m,p,g} \int_{-kn_2}^{kn_2}\!\!\!\!\! d\beta \, g_{\nu,e,g}g^*_{\nu,e',g} \vert_{\omega=\omega_{eg}}, \label{Eq::TotaleeDecayRate}
	\end{align}
and the small energy shift is absorbed into the transition frequency $\omega_{eg} = (E_e - E_g)/\hbar$.  
Equation (\ref{Eq::TotaleeDecayRate}) captures the modification of the spontaneous emission rate due to the nanofiber.  
The first sum describes decay into the guided modes and the second into the unguided radiation modes \cite{ nha_cavity_1997,klimov_spontaneous_2004,le_kien_spontaneous_2005,maslov_distribution_2006, scheel_directional_2015}. The decay rate of a given excited state into all guided modes is given by
	\begin{equation}
		\Gamma_e^{\oneD}\!\!=\!\! 2\pi\!\! \sum_{b,p,g}\!\! |g_{\mu,e,g} |^2_{\omega \!=\! \omega_{eg}} \!\!\!=\!\!  \frac{ 2\pi }{\hbar}\! \frac{ \omega_{eg} }{v_g}\!\! \sum_{b,p,g}\!\! \big|\bra{e}\hat{\mathbf{d}}\ket{g} \!\cdot\! \mathbf{u}_{bp}(\!\br'_\perp\!)\big|^2 \! .
	\end{equation}
This is in agreement with the expected expression from the guided-mode contribution to the dyadic Green's function in \erf{Eq::ImGreenLocal},
	\begin{equation} \label{Eq::Gamma1DGreens}
		\Gamma_e^{\oneD} =  \frac{2}{\hbar} \sum_{g}  \bra{g}\hat{\mathbf{d}}\ket{e}\!\cdot\! 
{\rm Im} \Big[\tensor{\mathbf{G}}^{(+)}_g(\br', \br'; \omega_{eg} ) \Big] \!\cdot\! \bra{e}\hat{\mathbf{d}}\ket{g},
	\end{equation}
which is enhanced over the free-space rate by the Purcell factor. 

Here we are interested in linear response for excitation far from resonance. We follow Ref. \cite{le_kien_propagation_2014} and consider an atom sufficiently far from the fiber surface such that the modification of the spontaneous emission rate is small.   In this case the decay rate is approximated as $\Gamma_{ee'} \approx \delta_{e,e'} \Gamma_{e}$, where $\Gamma_e$ is the total decay rate from excited state $\ket{e}$, given by the diagonal elements of \erf{Eq::TotaleeDecayRate}.  In steady state, the dipole operator in the linear regime ($\hat{\sigma}_{ee'} \rightarrow 0 $) is approximately
	\begin{align}
		\hat{\sigma}_{ge} &\!\approx\!  -\!\sum_{g'}\! \hat{\sigma}\!_{gg'}\!\!\int_0^{\infty}\!\!\!\mathrm{d}\omega \bigg(\!  \sum_{b,p}\!  
\frac{g_{\mu, e,g'}}{\omega\!-\!\omega_{eg} \!+\! i \Gamma\!_{e}/2  }\, \hat{a}_\mu (t_0) \nn\\
	&\!\! +\!\sum_{m,p}\! \int_{\!-kn_2}^{kn_2}\!\!\mathrm{d}\beta  \frac{g_{\nu, e,g'}}{\omega\!-\!\omega_{eg} \!+\! i \Gamma\!_{e}/2 } \,\hat{a}_\nu (t_0)  \!\bigg)e^{\!-i\omega (t\!-\!t_0)} . 
	\end{align}
By substituting this into \erf{Eq::aguidedEOM} and defining asymptotic modes, $\hat{a}^{\inp}(\omega) = \lim_{t_0\rightarrow -\infty} \hat{a}(t_0) e^{i\omega t_0}$, $\hat{a}^{\out}(\omega) = \lim_{t\rightarrow +\infty} \hat{a}(t) e^{i\omega t}$ \cite{fan_input-output_2010}, we obtain the input-output relationship for the guided modes, 
	\begin{align} \label{Eq::aout}
		\hat{a}^{\out}_\mu (\omega) &\!=\! \hat{a}^{\inp}_\mu (\omega) \!-\! 2\pi i\sum_{b',p'} 
\sum_{e,g,g'}\!\!\hat{\sigma}_{gg'}\frac{ g_{\mu,e,g}^* g_{\mu'\!\!,e,g'}}{ \omega \!-\! \omega_{eg} \!+\! i \Gamma\!_{e}/2 }\hat{a}^{\inp}_{\mu'}(\omega) \nn\\
&\!\!\!\!\!\!\!\!-\! 2\pi i\!\sum_{m\!,p} \sum_{e\!,g\!,g'}\! \int^{kn_2}_{-kn_2}\!\!\!\! \mathrm{d}\beta \,\hat{\sigma}\!_{gg'}\frac{ g_{\mu,e,g}^* g_{\nu'\!\!,e,g'}}{ \omega \!-\! \omega_{eg} \!\!+\! i \Gamma\!_{e}/2 }\hat{a}^{\inp}_{\nu}(\omega).
	\end{align}
	
This input-output relation contains the phase shift on forward scattered modes as well as attenuation due to elastic scattering into all other modes. For a probe with frequency $\omega_0$, \erf{Eq::aout} agrees with the expected form given by the Lippmann-Schwinger equation in the first Born approximation \footnote{In a careful derivation of the Lippman-Schwinger scattering equation, it is not the total Green's function that appears in \erf{Eq::ScatteredField} but rather a related dyadic quantity, $\mbf{K}(\br,\br', \omega_0) = \mbf{G}(\br,\br', \omega_0) + \delta(\br-\br')/n^2(\br)$ \cite{wubs_multiple-scattering_2004}. 
This function arises by proper accounting of the scatterer's coupling to the \emph{displacement} rather than the electric field \cite{yao_ultrahigh_2009} with a distinction that becomes important at the source point $\mbf{r} = \mbf{r}'$. However, for lossless dielectrics $n(\mathbf{r})$ is real and ${\rm Im}[\tensor{\mathbf{G}}(\br',\br'; \omega)] = {\rm Im}[\tensor{\mathbf{K}}(\br',\br'; \omega)]$ \cite{yao_-chip_2010}. },
	\begin{align} \label{Eq::IOScatteredField}
		\!\!\!\hat{\mathbf{E}}^{(+)}_{\out,g}&(\br, \omega_0)=\hat{\mathbf{E}}^{(+)}_{ \inp, g}(\br, \omega_0)\nn\\
		&\!\!\!\!\!\!\!\!\!+\tensor{\mathbf{G}}_g^{(+)}(\br,\br',\omega_0)\!\cdot\! \poltens \!\cdot\! \big[\hat{\mathbf{E}}^{(+)}_{\inp, g}(\br', \omega_0)\!+\!\hat{\mathbf{E}}^{(+)}_{\inp,r}(\br', \omega_0) \big],
	\end{align}
by noting that for the guided-mode dyadic Green's function given in \erf{Eq::GreensGuided},
	\begin{align}
		&\int d^2 \mbf{r}_\perp \, \mathbf{u}^*_{\mu} (\br_\perp)\cdot \tensor{\mathbf{G}}_g^{(+)}(\br,\br',\omega_0)\cdot \poltens \cdot \mathbf{u}_{\mu'} (\br'_\perp) \nn\\
		=& i \frac{ 2\pi \omega_0}{v_g} \mathbf{u}^*_{b,p} (\br'_\perp) \cdot \poltens \cdot \mathbf{u}_{b'\!\!, p'} (\br'_\perp). 
	\end{align}
Here, the atomic polarizability operator \cite{buhmann_casimir-polder_2004, deutsch_quantum_2010,kien_dynamical_2013}, is given by
	\begin{equation} \label{Eq::PolarizabilityOperator}
		\poltens = - \frac{1}{\hbar} \sum_{e,g,g'}\ket{g}\frac{\bra{g}\hat{\mathbf{d}}\ket{e}\bra{e} 
\hat{\mathbf{d}}\ket{g'}}{\Delta_{eg} + i \Gamma_{e}/2 }\bra{g'},
	\end{equation}	
and $\Delta_{eg} = \omega_0 - \omega_{eg}$ is the laser detuning from the atomic transition.	
For an atom in ground state $\ket{g}$ and polarization $p$, the phase shift can be expressed as \cite{le_kien_propagation_2014}
	\begin{align} \label{Eq::PhaseShiftMultilevel}
		\delta  \phi_{p,g} &=2\pi \frac{ \omega_{0}}{v_g} \mathbf{u}^*_{+, p}(\br'_\perp) \cdot {\rm Re}\big[\bra{g} 
\hat{\tensor{\boldsymbol{\alpha}}} \ket{g} \big] \cdot \mathbf{u}_{+, p}(\br'_\perp) \nn\\
&= - \frac{ \omega_{0} }{v_g} \sum_e \frac{2 \pi |\bra{e}\hat{\mathbf{d}}\ket{g} \cdot \mathbf{u}_{+,p}(\br'_\perp)|^2}{ \hbar  \Delta_{eg} } .
	\end{align}
%where $\Delta_{eg} = \omega_0 - \omega_{eg}$ is the laser detuning from the atomic transition frequency.  
We employ this dispersive response for QND measurement of atoms, as we describe in the next section.

%============== SECITON: QND Measurement ==============%
\section{QND measurement of atoms} \label{Sec::QNDMeasurement}

%========= FIGURE: Geometry (coupling strength, magic wavelength, area and detuning) =========%
\begin{figure}
\includegraphics[scale=1]{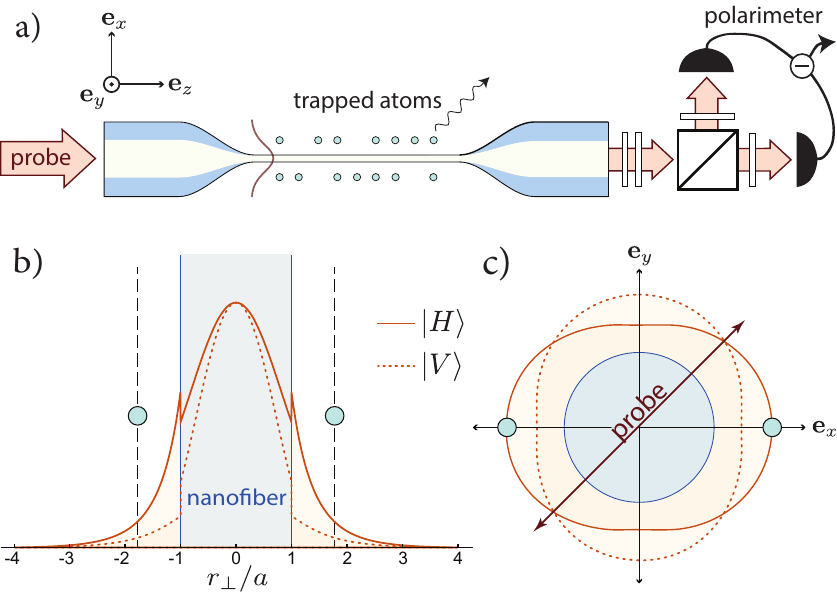}
\caption{Quantum interface for spin-polarization coupling of two 1D lattices of cold, trapped atoms and the guided modes of an optical nanofiber. a) Schematic of the interface.  A linearly polarized probe is launched into the nanofiber and the output light is analyzed in a polarimeter.  The atoms (green circles), trapped in the $x$-$z$ plane, couple to the evanescent portion of the guided $H$- and $V$-modes.  Contours of the $H$- and $V$-mode intensities in b) the $x$-direction and c) the transverse $x$-$y$ plane show the mode anisotropy at the atomic positions. }\label{Fig::Schematic}
\end{figure}
%=============================================

The dispersive interface between the atoms and nanofiber guided photons provides the entangling mechanism necessary to perform a QND measurement on the atoms.  
We restrict here to the quasilinear modes, $p =\{H,V\}$, of a single HE$_{11}$ guided mode at frequency $\omega_0$, whose form is given explicitly in \erf{Eq::QuasilinearModes}.  
In typical experimental configurations, two one-dimensional arrays of atoms are trapped on either side of the nanofiber, see Fig. \ref{Fig::Schematic}. 
We define coordinate axes $(x,y,z)$ with $z$ oriented along the fiber axis for forward propagation, and the two chains of atoms lie in the $x$-$z$ plane at azimuthal angles $\phi' = \{0, \pi\}$.
In the evanescent region, the $H$-mode is purely $\mathbf{e}_x$-polarized at $\phi = \pm \pi/2$ and the $V$-mode is purely $\mathbf{e}_y$-polarized at $\phi = \{0,\pi\}$.  
At other azimuthal angles the electric field is generally rotating along an ellipse in the $x$-$z$ plane.  The atoms at $\phi'=0$ experience $H$ and $V$ fields,
\begin{subequations}
	\begin{align}
		\mbf{u}_{b,H}(r_\perp, \phi = 0) = & \sqrt{2} \big[ \mathbf{e}_x u_r(r_\perp)+  i b \mathbf{e}_z  u_z(r_\perp) \big] \\
		\mbf{u}_{b,V}(r_\perp, \phi = 0) = & \sqrt{2} \mathbf{e}_y u_\phi(r_\perp), 
	\end{align}
\end{subequations}
where the real-valued functions $u_\alpha(r_\perp)$, given in \erf{Eq::ProfileFunctions}, depend only on the radial coordinate.  
On the opposite side of the fiber at $\phi' = \pi$, atoms experience the same transverse electric field, but the $z$-component changes sign.   This broken symmetry has been used to selectively address and separately control the two atomic arrays \cite{mitsch_exploiting_2014, mitsch_quantum_2014, sayrin_storage_2015}.  

We consider quasi-monochromatic fields at carrier frequency $\omega_0$ that are sufficiently narrowband, $\Delta \omega \ll \omega_0$. 
For each guided mode we define input propagating, continuous-mode field operators in the interaction picture \cite{gardiner_input_1985, blow_continuum_1990, le_kien_correlations_2008},
	\begin{align}
		\hat{a}_{b,p}(z,t) =\frac{1}{\sqrt{2 \pi}}  \int_0^{\infty}\!\!\!\!\! d \omega \, \hat{a}_{b,p}(\omega) e^{i[b \beta_0 z- (\omega-\omega_0) t ]}, 
	\end{align}
that satisfy the free-field commutation relations,
	\begin{equation} \label{Eq::InputOutputCommutation}
		\big[\hat{a}_{b,p}(z,t),\hat{a}^\dag_{b'\!,p'}(z'\!,t')\big]=\delta_{b,b'}\delta_{p,p'}  \delta(t\!-\!t'\!\!-\!(\!z\!-\!z')/v_g).
	\end{equation}
In terms of these propagating modes the quantized electric field operator, \erf{Eq::QuantizedElectricField}, becomes
	\begin{equation} \label{Eq::PropagatingElectricField}
		\!\!\!\hat{\mathbf{E}}^{(+)}(\!r\!_\perp\!,\phi,z;t\!) \!=\!\! \sum_{b,p}\!\! \sqrt{ \frac{2 \pi \hbar \omega_0}{ v_g} } \mathbf{u}_{b,p}(r\!_\perp\!,\phi) \hat{a}_{b,p}(\!z\!,t\!)  e^{i b \beta\!_0 z}\!,
	\end{equation}	
Considering here only the forward-propagating guided modes ($b=+$), we drop the $b$ index.  
The propagating electric field, \erf{Eq::PropagatingElectricField}, interacts with the trapped atoms via the dispersive light-shift Hamiltonian~\cite{deutsch_quantum_2010,kien_dynamical_2013,baragiola_open_2014},
	\begin{equation} \label{Eq::LightShiftHam}
		\hat{H}_{LS} = - \sum_{n=1}^{N_A} \hat{\mathbf{E}}^{(-)}(\mathbf{r}'_n ; t ) \cdot \poltens {}^{(n)} \cdot \hat{\mathbf{E}}^{(+)}(\mathbf{r}'_n ;t ),
	\end{equation}
where $\poltens {}^{(n)}$ is the atomic tensor polarizability operator, given in \erf{Eq::PolarizabilityOperator}, for the $n^{th}$ atom trapped near the nanofiber surface at position $\mathbf{r}'_n$.  We ignore here any effects of atomic motion and treat the atoms as localized at fixed positions in space.

The Lippmann-Schwinger scattering equation, \erf{Eq::IOScatteredField}, follows in the time domain as the evolution of coarse-grained input-ouput modes \cite{gardiner_input_1985, fan_input-output_2010, le_kien_propagation_2014}.  
Since multiple scattering is negligible and the propagation time across the ensemble is small compared to the atomic dynamics, we drop the position label $z$ and index the propagating fields by time alone; $\hat{a}_{b,p}(z,t) \rightarrow \hat{a}_{b,p}(t) $ as is standard in input-output theory \cite{gardiner_input_1985, stockton_deterministic_2004} . 
It follows that the effects of retardation can be ignored, in which case each term in the sum over atoms contributes equally for all atoms (for details see \cite{le_kien_correlations_2008, baragiola_open_2014}).  
The forward-propagating output fields are then given by the Fourier transform of \erf{Eq::aout}, yielding \cite{le_kien_correlations_2008} 
\begin{align} \label{Eq::ScatteringSolution}
		\hat{a}^{\rm out}_{p}(t) = \hat{a}^{\rm in}_{p}&(t) + i  \frac{2\pi \omega_0}{v_g}\sum_{p'}\Big[ N_0  \mathbf{u}^*_p(r'_\perp,0) \cdot \poltens \cdot  \mathbf{u}_{p'}(r'_\perp,0)  \nn\\
		&+ N_\pi \mathbf{u}^*_p(r'_\perp,\pi) \cdot \poltens \cdot  \mathbf{u}_{p'}(r'_\perp,\pi) \Big] \hat{a}^{\rm in}_{p'}(t),
	\end{align} 
where $\{N_0,N_\pi \}$ are the total number of atoms trapped at $\phi' = \{0,\pi\}$. The quantum effects from the first term give rise to shot noise in the transmitted field at the detector.  
The second term represents scattering into the guided modes, as described by the dyadic Green's function, \erf{Eq::Gamma1DGreens}.  

We introduce the vector Stokes operators that describe the polarization of the propagating fields in the quasilinear $HV$-basis,
\begin{subequations}\label{Eq::StokesComponents}
	\begin{align} 
		\hat{S}_1(t) & = \smallfrac{1}{2}\big[ \hat{a}^\dag_H(t) \hat{a}_H(t)-\hat{a}^\dag_V(t) \hat{a}_V(t) \big], \\
	 	\hat{S}_2(t) & = \smallfrac{1}{2}\big[ \hat{a}^\dag_H(t) \hat{a}_V(t)+\hat{a}^\dag_V(t) \hat{a}_H(t) \big], \\ 
		\hat{S}_3(t) & = \smallfrac{1}{2i}\big[ \hat{a}^\dag_H(t) \hat{a}_V(t) -\hat{a}^\dag_V(t) \hat{a}_H(t) \big],
	\end{align}
\end{subequations}
that satisfy equal-$z$ commutation relations following from \erf{Eq::InputOutputCommutation},
	\begin{equation} \label{Eq::StokesCommutation}
		\big[\hat{S}_i(t), \hat{S}_j(t')\big] =i \epsilon_{ijk} \delta(t-t')  \hat{S}_k(t).
	\end{equation}
These, along with the total photon flux operator,
	\begin{equation}
		\hat{S}_0(t) = \smallfrac{1}{2}\big[ \hat{a}^\dag_H(t) \hat{a}_H(t)+\hat{a}^\dag_V(t) \hat{a}_V(t) \big],
	\end{equation}
are used to reexpress the Hamiltonian, \erf{Eq::LightShiftHam}, in the $HV$-basis,
	\begin{align}  
		\hat{H}_{LS} 	=& - 2 \pi \hbar k_0 n_g \sum_{\phi'=0,\pi}N_{\phi'} \nn\\
		&\Big\{\,\,\, \big[ \polcomp_{HH}(\phi')+\polcomp_{VV}(\phi') \big] \hat{S}_0(t) \nn\\
		&+  \big[\polcomp_{HH}(\phi')  - \polcomp_{VV}(\phi')  \big] \hat{S}_1(t) \nonumber \\
&+ \big[\polcomp_{HV}(\phi') + \polcomp_{VH}(\phi')  \big] \hat{S}_2(t) \nn\\
&+ i  \big[ \polcomp_{H\!V}(\phi')-\polcomp_{V\!H}(\phi') \big]\hat{S}_3(t) \Big\}, \label{Eq::GenHamiltonian} 
	\end{align}
The atomic couplings to the $\{H,V\}$ modes,
	\begin{align} 
		\polcomp_{p p'}(\phi') & \equiv |\mathbf{u}^*_p(r'_\perp, \phi')||\mathbf{u}_{p'}(r'_\perp, \phi')| \, \hat{\alpha}_{p p'}(\phi') , 
	\end{align}
are determined by components of the quantum mechanical tensor operator weighted by the transverse mode functions at the atomic position, $\hat{\alpha}_{p p'}(\phi') = \mathbf{e}^*_{p'}(\phi') \cdot \poltens \cdot \mathbf{e}_{p}(\phi') $, whose classical analog appeared in \erf{Eq::PhaseShift}.

We explore a QND measurement of ${}^{133}$Cs atoms in the electronic ground state, $6S_{1/2}$, via polarization spectroscopy based on the collective atom-light coupling described by the dispersive light-shift Hamiltonian in \erf{Eq::GenHamiltonian}. Polarization transformations occur due to the tensor nature of the atomic response,
	\begin{align} \label{Eq::Polarizability}
		\poltens &=  \sum_{f,f'} \charpol \sum_{i,j} \hat{\tensor{\mbf{A}}}(f,f'),
	\end{align}
where the operator $\hat{\tensor{\mbf{A}}}(f,f') = \sum_{i,j} \hat{A}_{ij}(f,f')\mathbf{e}_i \otimes \mathbf{e}_j$ decomposes into irreducible components within each ground hyperfine multiplet $f$ for light detuned near excited multiplet $f'$,  
	\begin{align} \label{Eq::PolarizabilityIrrep}
		\hat{A}_{i\!j}(f\!,\!f')&\!=\!  C_{f\!f'}^{(0)} \delta_{i,j}\!+\! iC_{f\!f'}^{(1)}\epsilon_{i\!j\!k}\hat{f}\!_k \nn\\
		&\quad\!+\! C_{f\!f'}^{(2)} \Big[ \smallfrac{1}{2} ( \hat{f}\!_i\hat{f}\!_j \!+\!\hat{f}\!_j\hat{f}\!_i )\!-\!\smallfrac{1}{3} \hat{\mathbf{f}}\!\!\cdot\!\hat{\mathbf{f}} \delta_{i,j} \Big]. 
\end{align}
Here, $\charpol = -\frac{\sigma_0}{8\pi k_0}\frac{\Gamma }{\Delta_{ff'}+i\Gamma/2}$ is the characteristic dynamic polarizability where $\sigma_0 = 3 \lambda^2/2\pi$ is the resonant scattering cross section, $\hat{\mathbf{f}}$ is the atomic  spin operator in hyperfine multiplet $f$, and $C_{ff'}^{(K)}$ are coefficients for irreducible rank-$K$ components defined in \cite{deutsch_quantum_2010}. 

In addition to the atomic tensor response, the nanofiber geometry gives rise to unique features of polarization spectroscopy not present in free space.  The spatial anisotropy of the intensity for the quasilinearly polarized guided modes leads to unequal scattering of the $H$ and $V$ modes, producing \emph{intrinsic} birefringence even for a purely scalar atomic polarizability.  
In particular, atoms trapped on the quasi-$H$ axis leads to a phase delay of this mode relative to the fast quasi-$V$ axis. 
This birefringence was exploited by Dawkins {\em et al.} \cite{dawkins_dispersive_2011} as a mechanism for implementing a dispersive QND measurement of the number of atoms trapped around the nanofiber, as we treat in the next section.

	%=================== Atom number measurement =====================%
	\subsection{Dispersive atom number measurement} \label{Sec::AtomNumberMeasurement}

The anisotropy of the guided modes provides a mechanism for counting the number of atoms trapped around the nanofiber based on polarization spectroscopy.  
We consider $N_A$ atoms, each in a completely mixed hyperfine spin state. In this case the atomic polarizability tensor in \erf{Eq::PolarizabilityIrrep} reduces to $\langle \hat{A}_{ij}(f,f') \rangle = C_{ff'}^{(0)} \delta_{i,j}$, and the collective interaction is determined entirely by the the scalar (rank-0) terms.  
With the atoms trapped along the quasi-$H$ axis, while $\expt{ \polcomp_{HH} } \neq  \expt{ \polcomp_{VV} }$, the off-diagonal elements in \erf{Eq::GenHamiltonian} do not contribute to the Birefringent interaction we are interested in and actually vanish ($\expt{ \polcomp_{HV} } = \expt{ \polcomp_{VH} } =0$) when $x-$, $ y- $ or $ z- $axis is chosen as the quantization axis which includes the one close to the optimal choice of quantization axis for spin squeezing we will discuss in the next section.  
Atoms on either side of the nanofiber experience the same scalar light shift yielding from  \erf{Eq::GenHamiltonian} the Hamiltonian for QND measurement of atom number,
	\begin{align}
		\hat{H}_{N} &= -2\pi \hbar k_0 n_g \!\!\!\!\!\sum_{\phi' = \{0,\pi \}}\!\!\!\!\!\! N_{\phi'} \big[ \expt{ \polcomp_{HH}(\phi')}  - \expt{ \polcomp_{VV}(\phi')} \big] \hat{S}_1(t)  \nonumber \\
		& =  \hbar \chiN N_A \hat{S}_1(t).  \label{Eq::MixedHamiltonian}
	\end{align}	
This birefringent interaction induces a rotation of the Stokes vector  around the $S_1$-axis on the Poincar\'{e} sphere through an angle, 
	\begin{equation} \label{Eq::RotationAngle}
		\chiN = \frac{\sigma_0}{\Abir}  \sum_{f,f'}  C_{ff'}^{(0)} \frac{\Gamma}{2 \Delta_{ff'}},
	\end{equation}
characterized by an effective area, $\Abir^{\!-\!1} \!\equiv\! (n_g/2) \big( |\mathbf{u}_{H}(\!\br'\!\!_\perp\!)\!|^2 \!-\! |\mathbf{u}_{V}(\!\br'\!\!_\perp\!)| ^2 \big)$.   

Dawkins {\em et al.} \cite{dawkins_dispersive_2011} used this interaction to make a dispersive measurement of $N_A$ via birefringence polarimetry in the usual way: launching linearly polarized light at 45$^\circ$ to the quasi-$H$ axis, $\mbf{u}_\inp = (\mbf{u}_H + \mbf{u}_V )/\sqrt{2}$, and measuring the differential power between the guided right-and left-circularly polarized photons. 
Thus, the integrated measurement is described by the operator $\hat{\mathcal{M}} \equiv \int_0^T dt' \hat{S}^{\rm out}_3(t').$  The shot-noise variance of the polarimeter, $\shotnoise =  \chiN^2 \dot{N}_L T$ for integration time $T$,  determines the fundamental resolution of the polarimeter.  
The smallest detectable atom number using this dispersive measurement is thus, $\delta N_A \sim ( \chiN^2 \dot{N}_L T)^{-1/2}$ \cite{smith_faraday_2003}.  
In an ideal setting, $\delta N_A$ can always be reduced by increasing the integration time, but in practice this time is limited by atom loss. As a coarse approximation we take this time to be $T=\gamma_s^{-1}$, where $\gamma_s$ is the photon scattering rate in free space, and assume perfect quantum efficiency of the detectors.  
For detuning $\Delta$ large compared to the excited hyperfine splitting on the D1- or D2-line ($j' = 1/2$ or $3/2$), the unit-oscillator scattering rate is 
$\gamma_s =\frac{\sigma_0}{A_{\rm in}}\left(\frac{\Gamma}{2 \Delta}\right)^2 \dot{N}_L $, 
with effective area determined by the probe at the atomic position, \erf{Eq::AreaIn}.  
In this limit, the rotation angle $\chiN = C^{(0)}_{j'} (\sigma_0/\Abir)(\Gamma/2\Delta)$, \erf{Eq::RotationAngle}, yields a shot noise-limited atom number resolution, 
	\begin{align} \label{Eq::AtomNumberResolution}
		\delta N_A  &\sim \frac{1}{C^{(0)}_{j'}}\sqrt {\frac{\Abir^2}{A_{\rm in} \sigma_0}},
	\end{align}
where $ C^{(0)}_{j'}=\sum_{f,f'}C^{(0)}_{ff'}$ are the far-detuned, rank-$0$ coefficients on a $j \rightarrow j'$ transition \cite{deutsch_quantum_2010}.  
Using the parameters reported by Dawkins \emph{et al.}~\cite{dawkins_dispersive_2011}, we find the shot-noise limited minimum detectable atom-number $\delta N_A \sim 10$ for atoms trapped at $ 1.8a\sim 2.0a $ from the fiber axis with a D2-line probe light. 

In practice, loss and decoherence limit the atom-number resolution~\cite{dawkins_dispersive_2011, zhang_collective_2012}. 
The experimental implementation reported by Dawkins \emph{et al.}~\cite{dawkins_dispersive_2011} implies a resolution of a few tens of atoms for $ 200\sim 1000 $ trapped atoms.  
A similar experiment based on a two-color QND measurement in a nanofiber geometry was recently carried out by \emph{B\'{e}guin et al.} \cite{beguin_generation_2014} to squeeze the uncertainty in the number of trapped atoms. They achieved an atom number uncertainty of $\delta N_A = 8$ for $N_A\sim2500$ atoms, well below standard quantum limit, $\delta N_A=\sqrt{N_A}$.

	%===================QND spin squeezing=====================%
	\subsection{Collective spin squeezing via QND measurement}

The same birefringent interaction, \erf{Eq::GenHamiltonian}, can be utilized in a QND measurement to squeeze the projection noise of the collective atomic spin.  We consider squeezing of the uncertainty associated with the ``clock states" of cesium, $\ket{\uparrow} = \ket{6S_{1/2}, f=4, m_f=0}$ and $\ket{\downarrow} = \ket{6S_{1/2}, f=3, m_f=0}$, which define a pseudospin within each atom and associated Pauli operators $\{\hat{\sigma}_1, \hat{\sigma}_2, \hat{\sigma}_3\}$.  The quantum uncertainty in the collective pseudospin,
	\begin{align}
		\jz = \frac{1}{2} \sum_{n=1}^{N_A} \hat{\sigma}_3^{(n)},  
	\end{align}
fundamentally limits the precision of atomic clocks \cite{wineland_spin_1992}. For atoms prepared in a spin coherent state (SCS) the projection noise, $\varz \big|_{\scs} \!\!=\! N\!\!_A/\!4$, sets the standard quantum limit for spin measurements. A spin squeezed state (SSS) exhibits reduced fluctuations, $ \varz \big|_{\rm SSS}  \!\!<\! N\!\!_A/\!4$, due to negative pairwise correlations between the atoms~\cite{kitagawa_squeezed_1993}. Spin squeezing is typically quantified with the metrological squeezing parameter defined by Wineland \emph{et al.}~\cite{wineland_spin_1992},
	\begin{align} \label{Eq::SqueezingParameter}
		\xi^2 \equiv N_A \frac{ \varz }{ \expt{\hat{J}_{||}}^2 },
	\end{align}
where $\expt{\hat{J}_{||}}$ is the mean collective spin along the direction of spin polarization. 

The clock states are defined to have zero projection of angular momentum with respect to a bias magnetic field that defines a quantization axis, $\mathbf{e}_{\tilde{z}}$.  
Within the clock-state subspace the rank-1 vector light shift in the dispersive Hamiltonian, \erf{Eq::GenHamiltonian}, vanishes since $\bra{\uparrow}\hat{f}_k \ket{\uparrow} =\bra{\downarrow}\hat{f}_k \ket{\downarrow} = 0$ for any direction of the spin, $k$, and any quantization axis, $\mathbf{e}_{\tilde{z}}$. 
Furthermore, as shown below, atoms on either side of the nanofiber experience the same birefringent coupling. 
The resulting Hamiltonian, restricted to the clock subspace, couples the guided field of the nanofiber to the $J_3$-component of the collective pseudospin. The interaction has contributions from both the scalar and tensor light shifts,
	\begin{align} \label{Eq::ClockHamiltonian}
		\hat{H}_{J_3} = \hbar \Big\{ & \big[ \big( \chi_{H,\uparrow} +\chi_{V,\uparrow} \big) - \big( \chi_{H,\downarrow} + \chi_{V,\downarrow}\big) \big] \jz \hat{S}_0(t) \\
		+ & \big[  \big( \chi_{H, \uparrow} - \chi_{V,\uparrow} \big) - \big(\chi_{H,\downarrow} - \chi_{V,\downarrow} \big) \big]  \jz \hat{S}_1(t) \Big\}, \nonumber
	\end{align}
where the coupling strength between an atom in the clock subspace and a photon with polarization $p = \{H,V\}$ is
	\begin{equation} \label{Eq::ClockCouplingStrength}
		\chi_{p,f} \equiv - 2\pi k_0 n_g  | \mathbf{u}_p(\mbf{r}_\perp)|^2 \bra{f,0}\hat{\alpha}_{pp}  \ket{f,0},
	\end{equation}
and $f = \{4,3\}$ labels $\{\uparrow,\downarrow\}$.  
The diagonal terms in the polarizability tensor are the same for atoms at positions above and below the nanofiber, and thus all atoms contribute equally. 
In addition, a constant birefringence proportional to $ \hat{J}_0\hat{S}_1 $ is neglected here as it can be canceled with a compensating waveplate. 
Finally, the first term in \erf{Eq::ClockHamiltonian} does not affect polarization spectroscopy, but will act to rotate the pseudo-spin around the $J_3$-axis of the generalized Bloch sphere proportional to classical intensity fluctuations.
While this does not affect the squeezing of projection noise in $\hat{J}_3$, it affects the metrologically relevant squeezing by adding uncertainty to the direction of the mean spin.  By choosing a ``magic frequency" at which the light shifts on the two clock states are equal, 
	\begin{align} \label{Eq::MagicWavelengthCondition}
		\chi_{H,\uparrow} +\chi_{V,\uparrow}  = \chi_{H,\downarrow} + \chi_{V,\downarrow},
	\end{align}
this term can be canceled \cite{chaudhury_continuous_2006}, where we have ignored the imaginary part of the coupling strengths in the dispersive regime.
Using the D1-line of $^{133}$Cs atoms as the probe light, there are two magic-frequency solutions, $ \magic{3} $ and $\magic{4}$, shown in \frf{Fig::CouplingStrength}(a).  

Because the guided probe light at the position of the atom will generally be elliptical, the light-shift interaction coherently couples different magnetic sublevels in a given manifold $f$, and thus does not conserve $\hat{J}_3$.  For example, the ellipticity of the probe light leads to a fictitious magnetic field proportional to $i \mathbf{E}^{(-)}_{\inp}(\br') \times \mathbf{E}^{(+)}_{\inp}(\br')$ that causes a precession of the spin within hyperfine manifold $f$.  This can be mitigated by a sufficiently strong bias magnetic field compared to the fictitious field \cite{smith_continuous_2004}.

%========= FIGURE: Geometry (coupling strength, magic wavelength, area and detuning) =========%
\begin{figure}
\includegraphics[scale=0.44]{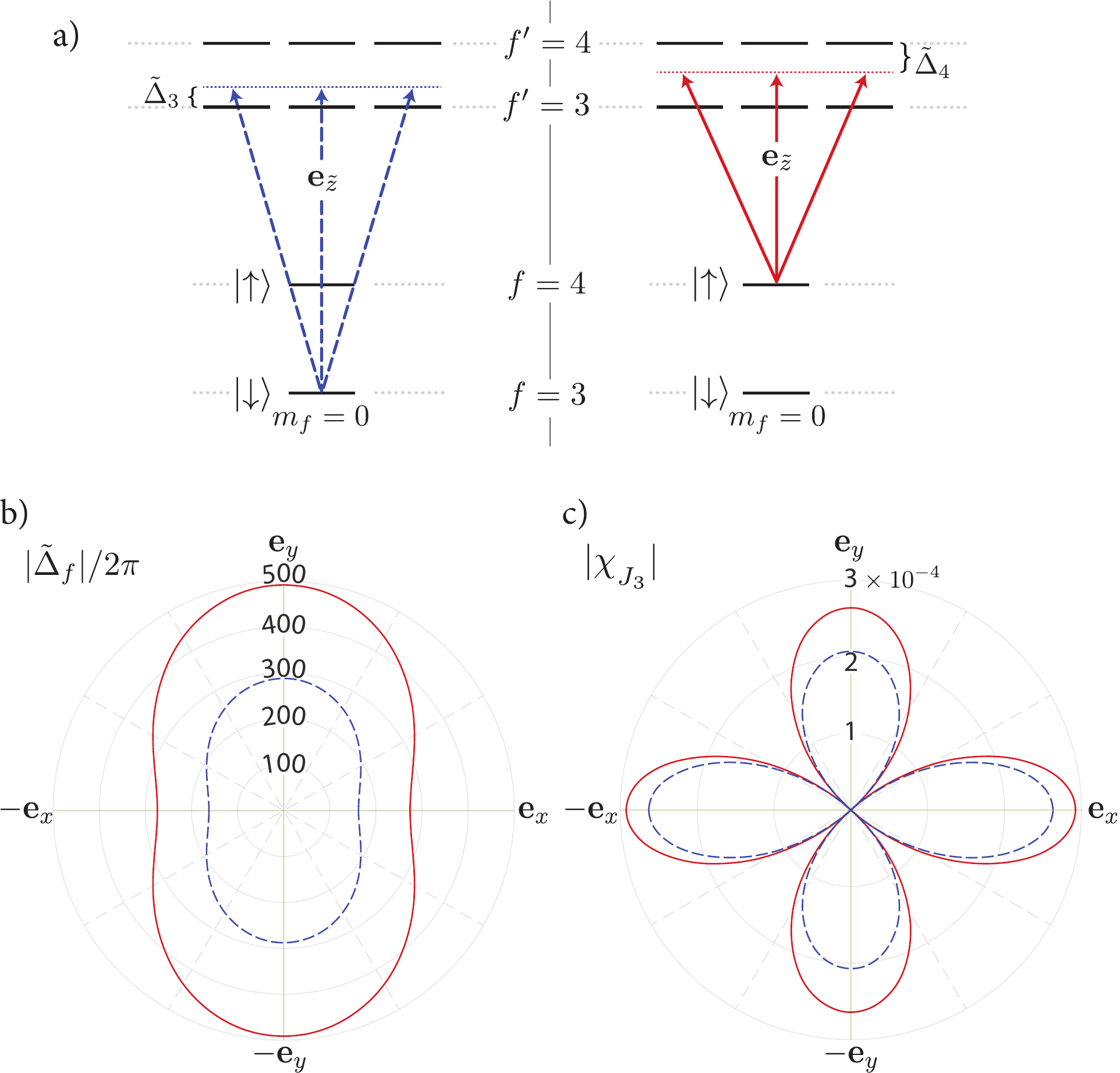}
\caption{Parameters of the atom-light interface using the clock states of $^{133}$Cs.  
(a) Energy level structure for atoms probed with one of two magic frequencies on the D1-line. 
(b) Magnitude of the magic detunings at which the clock states are equally light-shifted, $| \tilde{\Delta}_{f}|/2\pi \equiv | \omega_0 - \magic{f} |/2\pi$ (in units of MHz). There are two solutions shown as blue (dashed) and red lines. (c) The coupling strength, as measured by the magnitude of the polarization rotation angle on the Poincar\'{e} sphere, $|\chi_{_{ J_3}}|$, for an atom trapped in the $x$-$z$ plane at a distance $ r'\!_\perp=1.8a $ from the fiber center.  In both (b) and (c) we plot the parameter as the direction of the clock-state quantization axis is varied in the $x$-$y$ plane, for the two possible choices of magic detunings.
See text for details.}\label{Fig::CouplingStrength}
\end{figure}
%=============================================

The remaining QND interaction Hamiltonian is
	\begin{equation} \label{Eq::FaradayHam}
		\hat{H}_{J_3} = \hbar \chieff \jz \hat{S}_1(t),
	\end{equation}
where the rotation angle on the Poincar\'{e} sphere at the magic wavelength is,
\begin{equation}\label{eq:chiJ3}
\chieff = \big( \chi_{H, \uparrow} \!-\! \chi_{V,\uparrow} \big) \!-\! \big(\chi_{H,\downarrow} \!-\! \chi_{V,\downarrow} \big) = 2(\chi_{H, \uparrow}\!-\!\chi_{H, \downarrow}).
\end{equation}
In the standard way, squeezing the uncertainty in $\jz$ by QND measurement can be generated by preparing the atoms in a SCS along $\jx$, passing a probe prepared along $\hat{S}_2$ with photon flux $\dot{N}_L$, and continuously monitoring the $S_3$-component of the guided light in a polarimeter. The measurement strength,
	\begin{align} \label{Eq::MeasurementStrength}
		\kappa \equiv |\chieff|^2 \dot{N}_L, 
	\end{align}
quantifies the rate at which we squeeze projection noise, with $\chieff$ given in \erf{eq:chiJ3}. 
In the absence of any decoherence, such a QND measurement for integration time $T$ squeezes the initial uncertainty in $\jz$ according to $(\Delta J_3^2)_{\rm out}= (\Delta J_3^2)_{\rm in}/(1+r)$, where
	\begin{equation}
		r = \kappa T  (\Delta J_3^2)_{\inp}
	\end{equation}
is the integrated measurement strength \cite{hammerer_quantum_2010, baragiola_three-dimensional_2014}.

The strength of the birefringent interaction arises from two fundamental sources.  The anisotropy of the   $H$ and $V$ polarized modes leads to a polarization-dependent index of refraction, as described in Sec.~\ref{Sec::AtomNumberMeasurement}.  In addition there is a dependence of the atom-photon coupling on the internal spin state of the atom due to the atomic tensor polarizability.  In particular, we are interested in the dependence on the two clock states of the atom.  This spin-dependent coupling will depend on the choice of quantization axis that defines the clock state with projection $m_f=0$. 

We combine these two effects and obtain a compact expression for the coupling strength $\chieff$ using the irreducible tensor decomposition of the atomic polarizability, \erf{Eq::PolarizabilityIrrep}.  
Let $\{\mathbf{e}_{\tilde{x}},\mathbf{e}_{\tilde{y}}, \mathbf{e}_{\tilde{z}}\}$ be a space-fixed Cartesian coordinate system, where $\mathbf{e}_{\tilde{z}}$ defines the quantization axis of the atom, set by the magnetic field.  Because of the azimuthal symmetry of clock state around the $\qaxis$-axis, the polarizability tensor is diagonal in that basis.  
Noting that $\langle f,0 | \hat{f}_{\tilde{z}}| f,0 \rangle =0$ and $\langle f,0 | \hat{f}_{\tilde{x}}^2| f,0 \rangle = \langle f,0 | \hat{f}_{\tilde{y}}^2| f,0 \rangle = \langle f,0 | \hat{\mathbf{f}}^2| f,0 \rangle /2 =f(f+1)/2$, it follows that the expectation value of the irreducible rank-2 component of the atomic polarizability is
	\begin{align} \label{Eq::CouplingAngleMagic}
		\!\!\!\!\!\langle f,\! 0 | \poltens \phantom{}^{(\!2\!)}\!| f,\!0 \rangle \!=\!\! \sum_{f'}\! \charpol C^{(\!2\!)}_{f\!f'} \frac{f(f\!\!+\!\!1)}{6} \Big( \unittens \!\!-\! 3\qaxis \!\!\otimes\! \qaxis\! \Big).
	\end{align}
The combined scalar and tensor light shifts yield a coupling strength, \erf{Eq::ClockCouplingStrength},
	\begin{align}
		\chi_{p,f} &  = n_g \sigma_0 \big(  a_{f} \left|\mathbf{u}_p(\br'_\perp)\right|^2 - b_{f} \left|\mathbf{e}_{\tilde{z}} \cdot \mathbf{u}_p(\br'_\perp)\right|^2 \big), \label{Eq::ClockStateCoupling}
	\end{align}
with coefficients that depend on detunings and atomic structure,
	\begin{align}
		a_f &= \sum_{f'}  \Big(C^{(0)}_{ff'} + \frac{f(f+1)}{6} C^{(2)}_{ff'} \Big) \frac{\Gamma}{4 \Delta_{ff'}},\\
		b_f &= \frac{f(f+1)}{2}\sum_{f'} C^{(2)}_{ff'}  \frac{\Gamma}{4 \Delta_{ff'}}.
	\end{align}
At the magic wavelength set by \erf{Eq::MagicWavelengthCondition},
\begin{equation}
	\frac{a_4-a_3}{b_4-b_3} =  \frac{ |\mathbf{e}_{\tilde{z}} \cdot \mathbf{u}_V(\br'_\perp)|^2  + |\mathbf{e}_{\tilde{z}} \cdot \mathbf{u}_H(\br'_\perp) |^2 }{ |\mathbf{u}_H(\br'_\perp)|^2 + |\mathbf{u}_V(\br'_\perp)|^2 },
\end{equation}
which depends on the choice of quantization axis.

We write the effective rotation angle in the Hamiltonian, \erf{Eq::FaradayHam}, as
	\begin{align} \label{Eq::chieff}
		\chieff = \frac{\sigma_0}{A_{J_3}} \frac{\Gamma}{ 2 \Delta_{J_3}},
	\end{align}
with an ``effective detuning" set by the magic-wavelength condition,
	\begin{align} \label{Eq::SqueezingEffectiveDetuning}
		 \Delta_{J_3}^{-1} \equiv \frac{4}{\Gamma} (b_4 \!-\! b_3) =\!   \sum_{f'}\!  \left( C^{(2)}_{4f'}\!\frac{10}{\Delta_{4f'}} \!-\!  C^{(2)}_{3f'}\!\frac{6}{ \Delta_{3f'} } \right),
	\end{align}
and an effective area given by
	\begin{align} \label{Eq::SqueezingModeArea}
		\!\!\!\!\!A_{J_3}^{\!-\!1} & \!=\! n_g \frac{ |\mathbf{e}_{\tilde{z}} \!\cdot\! \mathbf{u}_V(\!\br'\!\!_\perp\!)\!|^2 |\!\mathbf{u}_H(\!\br'\!\!_\perp\!)\!|^2 \!-\! |\mathbf{e}_{\tilde{z}} \!\cdot\! \mathbf{u}_H(\!\br'\!\!_\perp\!) \!|^2 |\!\mathbf{u}_V(\!\br'\!\!_\perp\!)\!|^2 }{ |\mathbf{u}_H(\br'\!\!_\perp)|^2 + |\mathbf{u}_V(\br'\!\!_\perp)|^2 } \!.
	\end{align}	
We see here the explicit dependence of the coupling strength on both the anisotropy of the modes and on the tensor atomic response, which in turn depends on a particular choice of clock states.  The quantization axis that maximizes $\chieff$ is that which minimizes $A_{J_3}$ at a given magic detuning.  
Since the $z$-component of the guided modes is $90^\circ$ out-of-phase with the transverse components, the quantization axis maximizing the atom-light coupling is specified by an angle in the transverse $x$-$y$ plane, $\varphi$,
	\begin{align} \label{Eq::QuantizationAxis}
		\qaxis = \cos \varphi \mbf{e}_x + \sin \varphi \mbf{e}_y.
	\end{align}
The dependence of the magic detunings on the direction of quantization axis is shown in \frf{Fig::CouplingStrength}(b) for atoms trapped at a typical distance of $r_\perp'=1.8a$ on the $x$-axis.   
In typical operating regimes, the magic frequencies are hundreds of MHz from resonance with either excited state, placing the interaction in the off-resonant, dispersive regime.  
Using these magic detunings, in \frf{Fig::CouplingStrength}(c) we show the variation in $\chieff$ as a function of $\varphi$.  
This suggests that, based solely on the strength of the coherent interaction, the $x$-axis is the optimal quantization axis. 
As we will see in the next section, the optimal quantization axis is significantly modified when decoherence due to optimal pumping is included.

	%====== SUBSECTION: Including optical pumping ======%
	\subsection{Decoherence due to optical pumping}\label{sec:decoherence}
	
The treatment above considers an idealized QND interaction. 
The coupling of the atoms to the probe, however, will always lead to scattering of photons into modes other than the forward-scattered guided mode.  
This is accompanied by optical pumping that destroys the entanglement associated with spin squeezing.  In addition it reduces the metrologically useful signal.  
The maximum achievable metrologically relevant squeezing is determined by the balance of this decoherence with the QND measurement. 

We model this using a first-principles stochastic master equation description (SME)~\cite{jacobs_straightforward_2006, baragiola_three-dimensional_2014},
	\begin{align} \label{Eq::SME}
		d \hat{\rho} = s\sqrt{\frac{\kappa}{4}} \mathcal{H}[\hat{\rho}] dW + \frac{\kappa}{4} \mathcal{L}[\hat{\rho}] dt + \sum_n \mathcal{D}_n [\hat{\rho}] dt,
	\end{align}
where $s = {\rm sign}(\chieff)$ and $\hat{\rho}$ is the collective atomic state. 
The measurement strength $\kappa =|\chieff|^2 \dot{N}_L$ determines the rate of the spin squeezing in the absence of decoherence.  
The first two terms describe the QND measurement, where $dW$ is a stochastic Weiner increment satisfying $dW^2 = dt$. 
The conditional dynamics that result from the measurement are generated by the superoperator
	\begin{align}
		\mathcal{H}[\hat{\rho}] = \jz \hat{\rho} + \hat{\rho} \jz - 2 \expt{\jz} \hat{\rho},
	\end{align}
and the collective Lindblad map is
	\begin{align}
		\mathcal{L}[\hat{\rho}] = - \smallfrac{1}{2} \big( \hat{\rho}  \jz^2 + \jz^2 \hat{\rho} \big) + \jz \hat{\rho} \jz.
	\end{align}
The final term in \erf{Eq::SME} describes the effect of optical pumping acting locally on each atom along the nanofiber. 

The optical pumping map is governed by a standard master equation~\cite{deutsch_quantum_2010}.  
Restricting to the two-dimensional subspace associated with the clock states, the action on the $n^{th}$ atom is
	\begin{align}
		\mathcal{D}_n[\hat{\rho}] &=  \!\sum_{f=3,4}\!\! \Big\{ \!-\! \frac{\gamma_{f}}{2} \big[ \hat{\rho} (\op{f,\! 0}{f,\! 0})^{(n)} \!+\! (\op{f,\! 0}{f,\! 0})^{(n)}\hat{\rho} \big]\nn\\
		&+\!  \sum_{\tilde{f} =3,4}\!\!  \gamma_{f \!\rightarrow\! \tilde{f}}(\op{\tilde{f},\! 0}{f,\! 0})^{(n)} \hat{\rho}(\op{f,\! 0}{\tilde{f},\! 0})^{(n)} \Big\}.
	\end{align}
Here, $\gamma_{f}$ is the total rate of photon scattering by atoms in state $\ket{f,0}$ and  $\gamma_{f \rightarrow \tilde{f} }$ is the rate of optical pumping between the clock states, $\ket{f,0} \rightarrow \ket{\tilde{f},0}$ (see Appendix~\ref{Appendix::Rates}). Expressed in terms of Pauli operators on the clock-state pseudospin, the map acts as
	\begin{align} \label{Eq::OpticalPumpingMapSchr}
		\mathcal{D}_n [\hat{\rho}] 
				=& - \bigg[ \frac{2(\gammau+ \gammad) - \gammauu - \gammadd}{4} \bigg] \hat{\rho} \nn\\
				&- \frac{ \gammau - \gammad - \gammauu + \gammadd }{4} \big( \hat{\sigma}_3^{(n)} \hat{\rho}+ \hat{\rho} \hat{\sigma}_3^{(n)} \big) \nn \\
		&+ \frac{\gammauu+\gammadd}{4} \hat{\sigma}_ 3^{(n)}\hat{\rho} \hat{\sigma}_3^{(n)} \nn\\
		&+ \gammaud  \hat{\sigma}_-^{(n)} \hat{\rho} \hat{\sigma}_+^{(n)} + \gammadu  \hat{\sigma}_+^{(n)} \hat{\rho} \hat{\sigma}_- ^{(n)}.   
	\end{align} 
There are three important features of this map that are not typical in a QND measurement of ideal spin-$\half$ particles.  
First, the map is not trace preserving because atoms can be pumped out of the clock states. 
Second, unequal rates of optical pumping for $\ket{\uparrow}$ and $\ket{\downarrow}$ polarize the mean $\expt{\jz}$ towards a value different from that found in the QND measurement. 
Third, owing to the large ground hyperfine splitting, photons arising from optical pumping of $f \rightarrow \tilde{f}=3$ and $f \rightarrow \tilde{f}=4$ are distinguishable, thus these processes destroy coherences between $\ket{\uparrow}$ and $\ket{\downarrow}$. 

We calculate the squeezing parameter as a function of time based on the evolution of atomic correlation functions, where operators evolve according to the adjoint form of the SME in \erf{Eq::SME}.  The collective atomic variables obey the following stochastic equations of motion (see Appendix \ref{Appendix::OpticalPumping}),
\begin{subequations}\label{Eq::collectivedynamics}
	\begin{align} 
		&d N_C = -\gamma_{00} N_C dt + 2\gamma_{03} \expt{\hat{J}_3} dt \label{Eq::NA}\\
		&d \expt{\hat{J}_1}  = -\gamma_{11} \expt{\hat{J}_1} dt  \label{Eq::MeanSpinDecay} \\
		&d \expt{\hat{J}_3}  = s\sqrt{\kappa} \varz dW -\gamma_{33} \expt{\hat{J}_3}dt + \smallfrac{1}{2}\gamma_{30} N_C dt   \\
		&d \varz  = - \kappa \big(\varz\big)^2 dt- 2 \gamma_{33} \varz dt \nn\\
		&\quad\quad \!+\! \smallfrac{1}{4} \big( 2\gamma_{33}\!-\!\gamma_{00} \big) N_C dt \!+\! \smallfrac{1}{2} \big( \gamma_{03} \!-\! 2 \gamma_{30} \big) \expt{\hat{J}_3} dt,   \label{Eq::varJz} 
	\end{align}
\end{subequations}
where the decay and feeding rates are given in \erf{Eq::DecayRates}.
The total number of atoms in the clock-state subspace is given by $N_C$, which primarily decays at rate $\gamma_{00}$. 
The final term in \erf{Eq::varJz}, proportional to $\expt{\hat{J}_3}$, is typically negligible since in most applications $\expt{\hat{J}_3} \ll N_C$.  
We retain this small correction since unbalanced optical pumping acts to polarize the atoms and alters the rate of atom loss.  

\begin{figure*}
\includegraphics[scale=0.42]{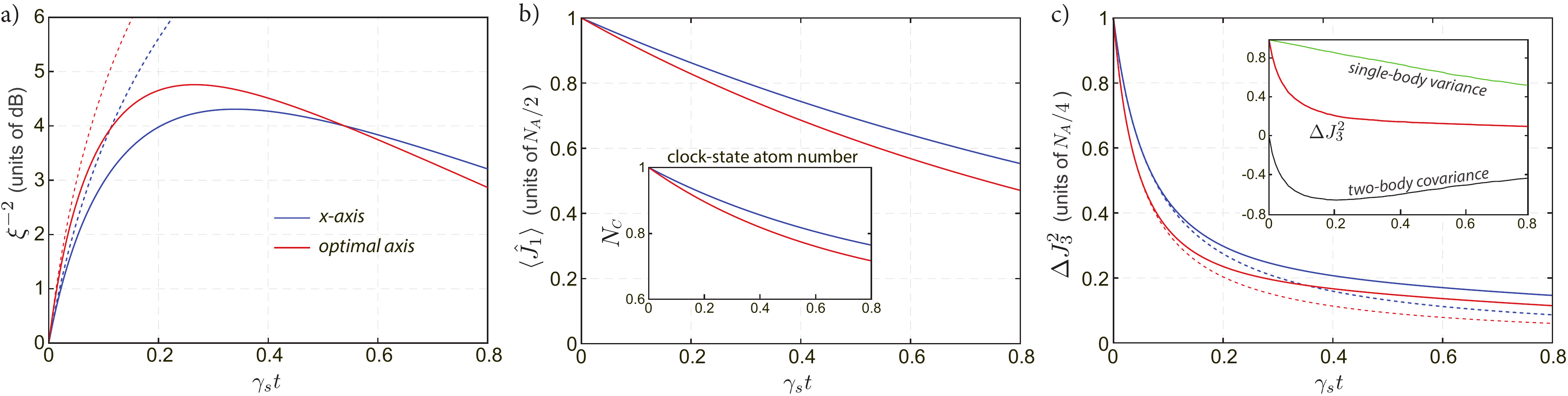}
\caption{Squeezing on the clock states as a function of time in units of the scattering rate $\gamma_s$ for $2500$ atoms trapped in the $x$-$z$ plane at a distance $ r'\!_\perp=1.8a$ from the axis of the nanofiber. The optimal quantization axis (red) is compared to the quantization along the $x$-axis (blue). 
Dashed lines indicate simulations without optical pumping; i.e. no decoherence. 
(a) Metrological spin squeezing parameter $\xi^{-2}$, \erf{Eq::SqueezingParameter}, in dB. 
(b) Collective mean spin $\expt{\hat{J}_1}$ and decaying atom number in the clock states, $N_C$ (inset).
(c) Conditional squeezed variance $\varz$. 
The inset shows the decomposition at the optimal quantization axis of $ \Delta J_3^2 $ (red dashed) into the single-body variance, $N\!\!_A (\!\Delta j_3^{(\!1\!)})^2$ (green) and the two-body covariance, $N\!\!_A(\!N\!\!_A\!\!-\!\!1\!)\expect{\!\Delta j_3^{(\!1\!)}\!\!\Delta j_3^{(\!2\!)}\!}$ (black), as given by \erf{Eq::VarianceDecomposition}.
}\label{Fig::Squeezing_Dynamics}
\end{figure*}

To find the peak squeezing in the presence of optical pumping, we numerically integrate Eqs. (\ref{Eq::NA}--\ref{Eq::varJz}) and then use \erf{Eq::SqueezingParameter} to calculate the metrological squeezing parameter, $\xi^2$, as a function of time. 
We choose here the magic frequency close to the $ f=4\leftrightarrow f'=4 $ transition, $ \magic{4} $, which is furthest from resonance with both excited hyperfine transitions. 
Typical time evolution is shown in \frf{Fig::Squeezing_Dynamics} for $2500$ atoms trapped a distance $r'\!_\perp=1.8a$ from the center of the nanofiber, where time is scaled to the characteristic scattering rate, 
\begin{align}\label{Eq::gamma_s}
\gamma_s \equiv \frac{\sigma_0}{A_{\rm in}}\frac{\Gamma^2}{4 \Delta_{J_3}^2} \dot{N}_L .
\end{align}  
We study the dynamics for two choices of quantization axis: (i) along the $x$-axis and (ii) along the numerically determined optimal axis.  Figure \ref{Fig::Squeezing_Dynamics}(a) shows the time evolution of the squeezing parameter.  We achieve a maximum squeezing of 4.7 dB when the clock states are chosen along the optimal axis; $\qangle_{\rm opt} \approx 86^\circ$ in \erf{Eq::QuantizationAxis}. 

%========= SQUEEZING DYNAMICS: \xi, variance, mean,  N_A =========
%================

The peak squeezing is ultimately limited by the combined effects of optical pumping on both $\expt{\jx}$ and $\varz$.  Here, as in a free-space model \cite{baragiola_three-dimensional_2014}, the primary factor that limits metrological squeezing is the decay of the collective mean spin $\expt{\jx}$. 
A scattered photon eliminates the initial coherence between $\ket{\uparrow}$ and $\ket{\downarrow}$ within a single atom, thus depolarizing $\expt{\jx}$.  
Atoms optically pumped to magnetic sublevels outside of the clock subspace decay $N_C$, further reducing $\expt{\jx}$. 
These effects are captured by the depolarization rate $\gamma_{11}$ in the equation for $\expt{\jx}$, \erf{Eq::MeanSpinDecay}, whose solution is plotted in \frf{Fig::Squeezing_Dynamics}(b).
  
We can gain deeper understanding in the microscopic effects of optical pumping on spin squeezing by looking at the evolution of the one and two-body correlation functions.  In terms of its constituent pseudospins, the collective variance takes the form
	\begin{align} \label{Eq::VarianceDecomposition}
	\varz & = N_A \big( \Delta j_3^{(1)} \big)^2 + N_A(N_A-1) \expt{ \Delta j_3^{(1)} \Delta j_3^{(2)} }
	\end{align}
for permutationally symmetric states considered here, where $(1)$ and $(2)$ label any two atoms in the ensemble. Loss of atoms affects the first (single-body) variance term, which scales as $N_A$.
The two-body correlations which contribute as $N_A^2$ to the collective fluctuations, 
	\begin{align}
		\expt{ \Delta j_3^{(1)} \Delta j_3^{(2)} } &\equiv \smallfrac{1}{4} \big( \expt{ \hat{\sigma}_3^{(1)} \otimes \hat{\sigma}_3^{(2)}  } - \expt{\hat{\sigma}_3^{(1)} }^2  \big).
\end{align}
have a much larger influence on the total variance. Spin-spin correlations at the heart of spin squeezing, $\expt{ \hat{\sigma}_3^{(1)} \otimes \hat{\sigma}_3^{(2)} }$, rapidly generated by the measurement backaction decohere by optical pumping according to Eqs. (\ref{Eq::TwoBodyDecay}--\ref{Eq::OperatorMap}),
	\begin{align} \label{Eq::TwoBodySpinDecay}
		\!\!\!\!\!\!\frac{d}{dt}\! \expt{\hat{\sigma}\!_3^{(\!1\!)}\!\! \otimes\! \hat{\sigma}\!_3^{(\!2\!)} }  \!\Big|_{\rm op}\!\!\!\! =\! - 2 \gamma_{33} \! \expt{\hat{\sigma}\!_3^{(\!1\!)} \!\!\otimes\! \hat{\sigma}\!_3^{(\!2\!)} } \!\!+\! \gamma_{30}\! \expt{ \hat{\mathbbm{1}}\!^{(\!1\!)}_C \!\!\otimes\! \hat{\sigma}\!_3^{(\!2\!)\!} \!\!+\! \hat{\sigma}\!_3^{(\!1\!)} \!\!\!\otimes\!\! \hat{\mathbbm{1}}\!^{(\!2\!)}_C },
	\end{align}
where $\hat{\mathbbm{1}}^{(n)}_C \equiv \big( \op{\uparrow}{\uparrow} + \op{\downarrow}{\downarrow} \big){}^{(n)}$ is the single-body projector onto the clock states.   
In addition, atoms that return to the clock subspace after scattering a photon inject additional noise into $\varz$.  
All of these effects are included in the equation for $\varz$, \erf{Eq::varJz}, whose overall and decomposed dynamical evolutions are shown in Fig. \ref{Fig::Squeezing_Dynamics}(c).

%========= FIGURE: Peak squeezing (\xi, measurement strength, scattering rates) =========
\begin{figure}
\includegraphics[scale=0.37]{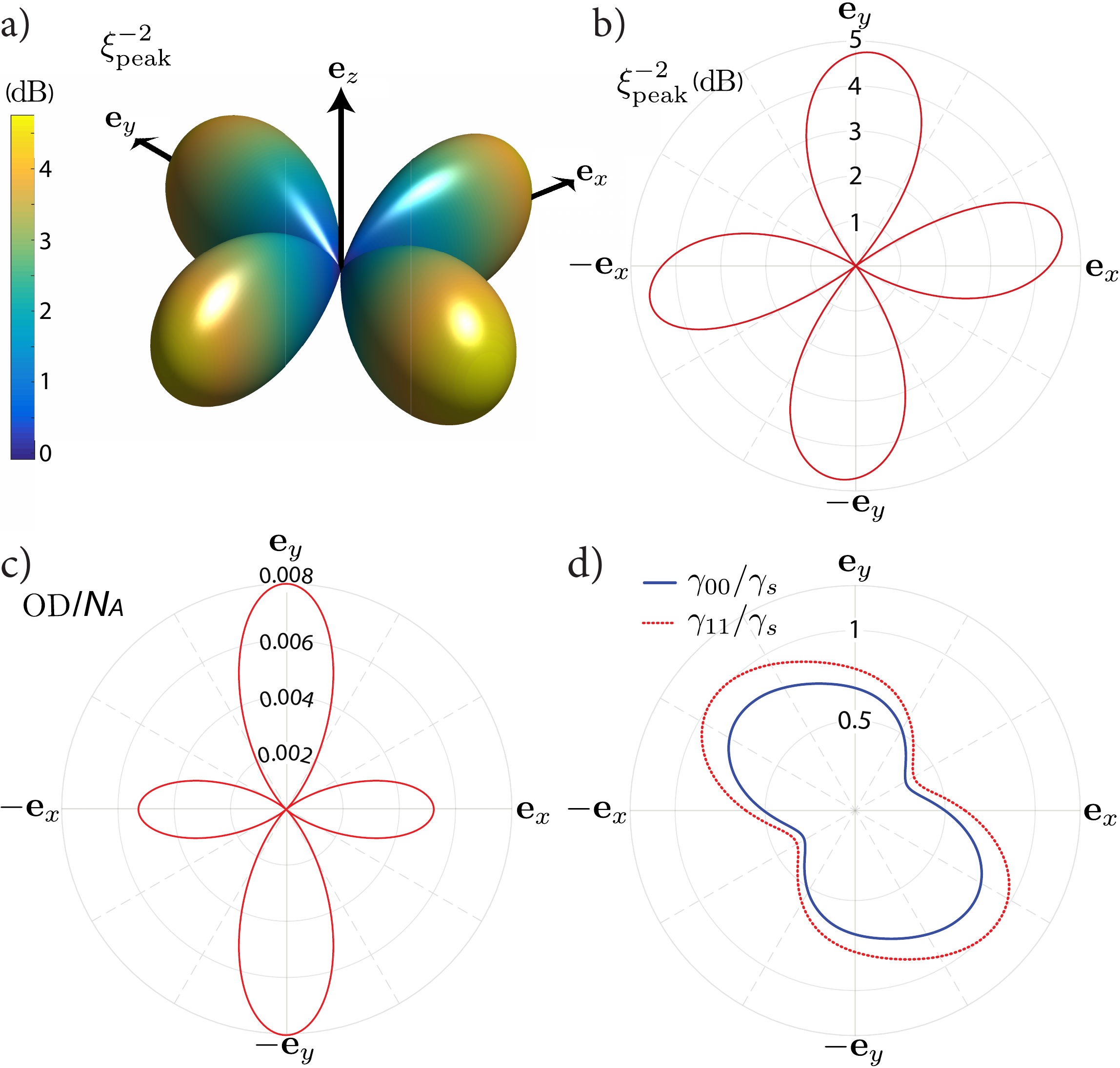}
\caption{Dependence of the parameters that determine metrological squeezing  on the direction of the quantization axis that defines the clock states, $\qaxis$.  In all cases we consider $2500$ atoms trapped in the $x$-$z$ plane at distance $ r'\!_\perp=1.8a$ from the axis of the nanofiber. 
In (b--d) $\qaxis$ is confined to the $x$-$y$ plane, where the optimal peak squeezing occurs.
(a,b)  Peak achievable squeezing at the maximum time, measured in dB, as a function of the direction of  $\qaxis$. 
(c) OD/$N_A$=$\sigma_0 A_{in}/A^2_{J_3}$, \erf{Eq::OD/atom}.
(d) Rates of atom loss, $\gamma_{00}$, and depolarization, $\gamma_{11}$ relative to the characteristic scattering rate $\gamma_s$.  See Eqs.~\eqref{Eq::lrate} and~\eqref{Eq::frate}. }\label{Fig::Squeezing_QuantizationAxis}
\end{figure}
%================

With our model, we explore optimal conditions for generating spin squeezing.   The choice of quantization axis $\qaxis$ that defines clock states affects both the measurement strength and the relative rates of optical pumping. We plot the peak squeezing as a function of the direction of $\qaxis$ in the $x$-$y$ plane in \frf{Fig::Squeezing_QuantizationAxis}(b). 
We gain insight into the tradeoffs between QND entangling interaction and decoherence by independent inspection of the measurement strength and optical pumping rates.  
First, the rate of squeezing is determined by the effective optical density per atom on resonance,
\begin{align}\label{Eq::OD/atom}
\mathrm{OD}/N_A \equiv \frac{\kappa}{\gamma_s}=\frac{\sigma_0 A_{\rm in}}{A_{J_3}^2},
\end{align} 
which peaks when $\qaxis$ is along the $y$-axis, as seen in \frf{Fig::Squeezing_QuantizationAxis}(c). 
Choosing $\qaxis$ along $y$, the OD/$N_A$ is about 50\% larger than along $x$-axis.  
The various forms of decoherence similarly vary with quantization axis, as seen in \frf{Fig::Squeezing_QuantizationAxis}(d), where we plot the dominant rate of atom loss, $\gamma_{00}$, and the depolarization rate of the mean pseudospin $\expt{\jx}$, $\gamma_{11}$. 
Because the magic frequency $\magic{4}$ is nearly equidistant from $f'=3$ and $f'=4$ when the quantization  axis is near the $y$-axis (see \frf{Fig::CouplingStrength}(b)) this choice of quantization axis provides more protection from decoherence.   
While the decoherence rates in \frf{Fig::Squeezing_QuantizationAxis}(d) are largest near the $y$-axis, the increase in $\kappa$ more than compensates to provide optimal peak squeezing.

Finally, we explore the optimal conditions as a function of the trapping geometry.  
The dispersive entangling interaction is based on the collective atomic coupling to the evanescent guided-mode fields, which decay exponentially away from the nanofiber surface, as seen in OD/$N_A$ plotted in \frf{Fig::Squeezing_Distance}(a). 
From Eq.~\eqref{Eq::varJz}, the optimal choice of quantization axis depends not only on distance from the fiber but also weakly on the atom number, \frf{Fig::Squeezing_Distance}(b) because of the competition between squeezing and decoherence.  
At the optimal quantization axis, the strong dependence of peak achievable squeezing on distance from the fiber is as seen in \frf{Fig::Squeezing_Distance}(c) along with the expected increase as more atoms contribute to the atom-light interface.  

Several effects limit the reliability of the simulations for atoms trapped very near the fiber surface as $r'\!\!_\perp\!\! \rightarrow\! a$. 
First, strong van der Waals interactions modify the light shifts and magic frequencies \cite{vetsch_eugen_optical_2010, lacroute_state-insensitive_2012}.  
Second, the optical pumping model used here breaks down when the local density of states is significantly modified by the presence of the dielectric nanofiber \cite{le_kien_spontaneous_2005, le_kien_scattering_2006}.   
At distances $r'\!\!_\perp \!>\! 1.5a$ the atoms' local environment is roughly that of unmodified vacuum \cite{le_kien_spontaneous_2005} and a free-space optical pumping model in \erf{Eq::OpticalPumpingMapSchr} suffices.

%======= FIG: Scaling with atom number and distance =======
\begin{figure*}\includegraphics[scale=0.4]{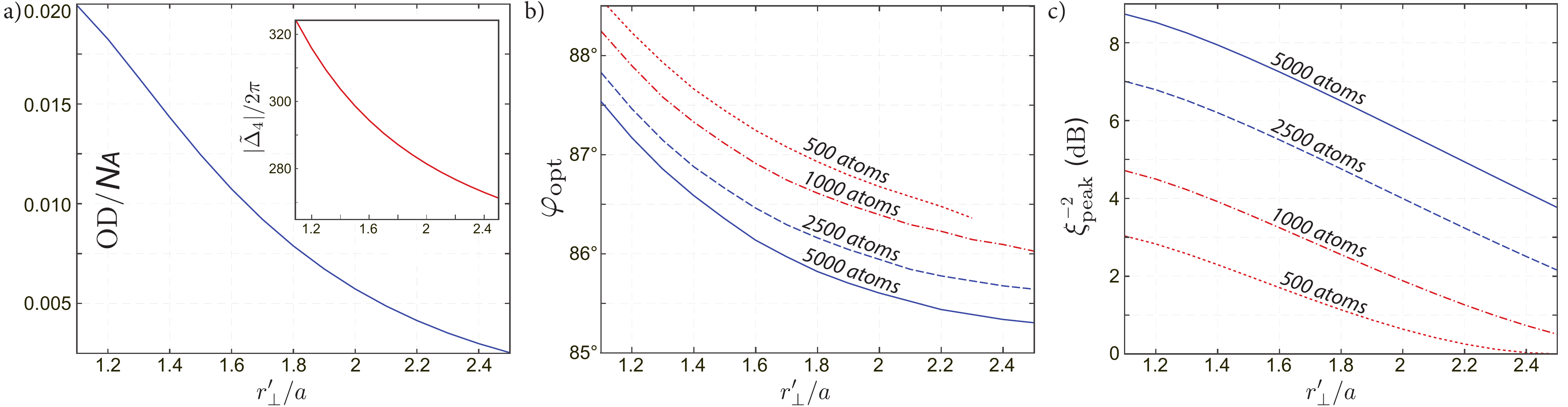}
\caption{Parameters that define squeezing as a function of trapping distance $r'_\perp$ and initial atom number $N_A$. (a) OD/$N_A$. Inset is the corresponding magic detuning in units of MHz (see text).
(b) Optimal quantization axis orientation angle $\qaxis$ in the $x$-$y$ plane for different atom numbers. 
The line with $ 500 $ atoms terminates when the squeezing effect is too weak to be observed ($ r'\!_\perp>2.3a $).
(c) Peak metrological squeezing at the optimal $\qaxis$ for different atom numbers.} \label{Fig::Squeezing_Distance}
\end{figure*}
%================================

%====== SECTION: Summary and outlook ======%
\section{Summary and Outlook} \label{Sec::Conclusion}

We studied the strong cooperativity in the atom-light interface that can be achieved based on atoms trapped in the evanescent field surrounding an optical nanofiber, and interacting with a guided mode in the dispersive regime. 
The key parameter that determines the coupling is the resonant optical density per atom. 
Due to the tight confinement of the guided mode over the entire chain of atoms this parameter is $ \mathrm{OD}/N_A\sim 10^{-2} $ for typical geometries used in current experiments, which approaches that achieved for atomic ensembles trapped inside optical cavities of moderate finesse \cite{chen_conditional_2011, zhang_collective_2012}.  
In contrast, the atom-light coupling for atoms in free space is typically orders of magnitude smaller, $ \mathrm{OD}/N_A \sim 10^{-6}$.  
Under ideal conditions the atom-light interaction is entirely symmetric along the nanofiber, providing a platform for long-range correlations independent of distance between the atoms. 
As the light is entirely guided, fiber- or waveguide-coupled atomic ensembles can be networked together or coupled to other physical systems in a hybrid platform \cite{hafezi_atomic_2012, liebermeister_tapered_2014, Meng2015nanowaveguide, Tiecke2015Efficient} for truly long-range entanglement generation and distribution. 

We calculated the dispersive response based on a modal decomposition of the dyadic Green's function, which provides a general method to calculate the induced phase shifts and polarization rotations of the guided modes. 
With this we studied the QND measurement of atoms via polarization spectroscopy. 
In particular, we studied squeezing of the collective pseudospin associated the atomic clock states of cesium. 
The atoms induce a birefringent index of refraction on the light, conditional on the spin state, which provides a mechanism for measuring the atomic spin projection and thus squeezing its uncertainty.  
Based on our formalism we calculated the nanofiber-enhanced measurement strength that determines the rate of squeezing.  

The peak squeezing one can generate depends on a detailed balance between the reduction of spin projection noise based on QND measurement and the damage done to the spin ensemble due to optical pumping.  
Both measurement and optical pumping arise from the same physical mechanism -- scattering of photons by atoms.  
The former corresponds to cooperative forward scattering into the guided mode whereas the latter corresponds to local scattering into all other modes, primarily the unguided ``radiation" modes.   The cooperativity, specified by the effective OD/$N_A$, determines the ratio of these two effects and thus the ultimate power of the quantum atom-light interface. 

We studied QND measurement using a first-principles stochastic master equation model, which allowed us to track the atomic correlation functions that define the metrologically relevant squeezing parameter.  
These include the atomic projection noise uncertainty as well as the length of the collective spin vector that defines the metrological signal.  We find that decoherence acts primarily to depolarize the mean pseudospin and optically pump atoms out of the clock subspace, which we treat as loss.  In addition, optical pumping decoheres the spin correlations at the heart of spin squeezing, but at a reduced rate compared with the effect on the mean pseudospin.  
The combined effect of QND measurement and decoherence yields a peak squeezing approaching 5 dB with $\sim 2500$ atoms. Larger enhancements in atom-light coupling and QND squeezing are possible with modest increases in the number of trapped atoms and/or for atoms trapped closer to the nanofiber surface.  

Whereas we have assumed here that atoms can be prepared in a desired clock state defined by a particular quantization axis, in practice such preparation will require optical pumping that may be challenging for atoms near the surface of the nanofiber.   In addition, though we have treated the atoms as localized at well defined points, in practice the atoms' thermal motion can reduce the strong coupling described here.  Our formalism  provides a starting point for developing models necessary to study the dynamics of optical pumping, including the possibility of cooling atoms to the vibrational ground state, where thermal motion is negligible. 

Finally, though we have treated here the case of strong coupling due solely to tight confinement of the guided mode for atoms near the surface of the nanofiber,  we can achieve even greater enhancement  by combining this effect with longitudinal confinement provided by fiber-based optical cavities~\cite{le_kien_intracavity_2009, wuttke_nanofiber_2012, yalla_cavity_2014, bohnet_reduced_2014, nayak_optical_2014}.  
The coupling can be further improved under EIT conditions that substantially slow the group velocity \cite{gouraud_demonstration_2015, sayrin_storage_2015, kumar_autler-townes_2015, le_kien_electromagnetically_2015}.  In addition, quantum control of the internal hyperfine state \cite{smith_quantum_2013-1} can greatly enhance the entangling power of the atom-light interface~\cite{trail_strongly_2010, norris_enhanced_2012}. 
For large enough coupling, QND measurement should allow production of highly entangled spin states beyond the Gaussian regime \cite{stockton_deterministic_2004, mcconnell_entanglement_2015}.

ACKNOWLEDGMENTS 
We gratefully acknowledge Sudhakar Prasad for helpful discussions regarding the dyadic Green's function. 
We thank the UNM Center for Advanced Research Computing for computational resources used in this work. 
This work was support by the AFOSR, under grant Y600242, and the NSF, under grants PHY-1212445, 1307520. 
\bibliography{Nanofiber}

%=========== APPENDIX ===========%
\begin{appendix}

%=========== APPENDIX: Nanofiber mode functions ===========%
\section{Guided-mode functions for the optical nanofiber} \label{Appendix::ModeFunctions}

In this Appendix we provide, for reference, the fundamental HE$_{11}$ solutions to the homogeneous wave equation, \erf{Eq::WaveEquationSource} with $\tensor{\boldsymbol{\alpha}} = 0$, for a cylindrical nanofiber of radius $a$ and index of refraction given by \erf{Eq::IndexofRefraction}.  At a given frequency, $\omega_0 = c k_0$, the magnitudes of the longitudinal and transverse wave vectors for a guided mode are related by $n^2 k_0^2 = \beta_0^2 + k_\perp^2$.  
The positive propagation constant, $\beta_0 \equiv \beta(\omega_0)$, is determined from the eigenvalue equation that results from enforcing physical boundary conditions at the fiber surface \cite{snyder_optical_1983},
	\begin{align}
		\!\!\!& \frac{J_0(ha)}{ha J_1(ha)} = - \frac{n_1^2+n_2^2}{2n_1^2} \frac{K'(qa)}{qa K_1(qa)} + \frac{1}{h^2 a^2} \nn\\
		\!\!\!\!\!\!\!\!&\, - \bigg[\! \bigg(\!\frac{n_1^2 \!-\! n_2^2}{2 n_1^2}\! \frac{K'\!(qa)}{qa K_1\!(qa)} \!\bigg)^2  \!\!\!+\! \frac{\beta_0^2}{n^2_1 k^2}\!\! \left(\!\frac{1}{q^2a^2} \!+\! \frac{1}{h^2a^2}\! \right)^2\! \bigg]^{\frac{1}{2}}\!.
	\end{align}
Inside the nanofiber the transverse wavevector is real, $k_\perp = q$, where $q=\sqrt{\beta_0^2- n_2^2k_0^2}$, and outside the nanofiber it is purely imaginary, $k_\perp = i h$, where $h=\sqrt{n_1^2 k_0^2 - \beta_0^2}$.  The vector eigenfunctions are expressed as $\mbf{f}_{\mu}(\br) = (2\pi)^{-1/2}\mbf{u}_{b,p}(\mbf{r}_\perp) e^{i b \beta_0 z}$, where the modes are indexed by frequency $\omega_0$, propagation direction $b = \pm$, and polarization $p$.

A relatively simple form for the guided-mode functions can be expressed in a cylindrical basis $(r_\perp, \phi, z)$ with longitudinal unit vector $\mathbf{e}_z$, oriented along the fiber axis.  
The transverse unit vectors are related to their fixed Cartesian counterparts via the relations
\begin{subequations}
	\begin{align}
		\mathbf{e}_{r_{\!\perp}}     &= \mathbf{e}_x \cos \phi + \mathbf{e}_y \sin \phi, \\
		\mathbf{e}_\phi &= - \mathbf{e}_x \sin \phi + \mathbf{e}_y \cos \phi.
	\end{align}
\end{subequations}
The transverse profile for the quasicircular guided modes, $p = \pm$, is
	\begin{align} \label{Eq::QuasicircularModes}
		\!\!\!\!\!\mbf{u}_{b,\pm}(\mathbf{r}\!_\perp) \!=\! \big[\mathbf{e}_{r_{\!\perp}}\!\! u_{r_{\!\perp}}\!\!(r\!_\perp) \!\pm\! i \mathbf{e}_\phi u_\phi(r\!_\perp) \!+\!  i b \mathbf{e}_z  u_z(r\!_\perp) \big]e^{ \pm i \phi}, 
	\end{align}
and for the quasilinear guided modes, $p = \{H,V\}$, is
	\begin{subequations} \label{Eq::QuasilinearModes}
	\begin{align}
		\mbf{u}_{b,H}(\mathbf{r}\!_\perp) = & \sqrt{2} \big[ \mathbf{e}_{r_{\!\perp}} u_{r_{\!\perp}}(r\!_\perp) \cos \phi \nn\\
		&-\! \mathbf{e}_\phi u_\phi(r\!_\perp) \sin \phi \!+\!  ib \mathbf{e}_z  u_z(r\!_\perp) \cos \phi \big] \\
		\mbf{u}_{b,V}(\mathbf{r}_\perp) = & \sqrt{2} \big[ \mathbf{e}_{r_{\!\perp}} u_{r_{\!\perp}}(r_\perp) \sin \phi \nn\\
		&+\! \mathbf{e}_\phi u_\phi(r_\perp) \cos \phi \!+\!  ib \mathbf{e}_z  u_z(r_\perp) \sin \phi \big]. 
	\end{align}
	\end{subequations}
The modes are expressed in terms of real-valued functions that depend only on the radial coordinate $r_\perp$,
	\begin{subequations} \label{Eq::ProfileFunctions}
	\begin{align} 
		u_{r_{\!\perp}}(r_\perp) =& u_0 \big[ (1-s) K_0(q{r_{\!\perp}}) + (1+s)K_2(q{r_{\!\perp}})\big] \\
		u_\phi(r_\perp) =& u_0\big[ (1-s) K_0(q{r_{\!\perp}}) - (1+s)K_2(q{r_{\!\perp}})\big] \\
		u_z(r_\perp) =& u_0 \frac{2 q}{\beta_0} \frac{K_1(qa)}{J_1(ha)} J_1(h{r_{\!\perp}}), \label{Eq::zprofile}
	\end{align}
	\end{subequations}
where $u_0$ is set by the normalization condition, $\int d^2 \mathbf{r}_\perp n(r_\perp) | \mathbf{u}_\mu(\br_\perp)|^2=1$, $J_n$ and $K_n$ are the $n^{th}$ Bessel functions of the first and second kind, $f'(x)$ indicates a derivative with respect to the argument $x$, and 
	\begin{align}
		s = \frac{1/(q^2 a^2)^{2} + 1/(h^2 a^2)^{2}}{[J'_1(ha)/haJ_1(ha) + K'_1(qa)/qaK_1(qa)]}.
	\end{align}  
Of particular interest is the $z$-component, \erf{Eq::zprofile}, which can become appreciable.  Note that the phase convention in Eqs. (\ref{Eq::QuasicircularModes}-\ref{Eq::ProfileFunctions}) has been chosen to emphasize properties of the quasilinear modes and differs from that of \emph{Le Kien et al.} -- for instance in Ref. \cite{le_kien_propagation_2014}.  
Further details about the guided-mode fields inside the nanofiber ($r_\perp\leq a$), the radiation (unguided) modes, and the quantized form of both can be found in Refs. \cite{sondergaard_general_2001, tong_single-mode_2004, kien_field_2004, le_kien_spontaneous_2005, vetsch_eugen_optical_2010}.

%===================APPENDIX: Photon scattering and optical pumping rates =====================%
\section{Photon scattering and optical pumping rates} \label{Appendix::Rates}	

In this Appendix we give the explicit expressions for the photon scattering rates used in Sec.~\ref{Sec::QNDMeasurement} following the formalism given in~\cite{deutsch_quantum_2010}.  The total rate of photon scattering by an atom in the clock state $\ket{f,0}$ is
	\begin{equation}\label{Eq::gammaf}
		\gamma_{f}=- \frac{2}{\hbar} {\rm Im} \big[ \bra{f,0} \hat{h}_{\rm eff}\ket{f,0} \big] ,
	\end{equation}
where the effective non-Hermitian light-shift Hamiltonian for one atom is 
\begin{align}
\hat{h}_{\rm eff} = - \hat{\mathbf{E}}^{(-)}_{\rm in}(\mathbf{r}' ; t ) \cdot \poltens \cdot \hat{\mathbf{E}}^{(+)}_{\rm in}(\mathbf{r}' ;t )
\end{align}
as follows from \erf{Eq::LightShiftHam}, where $\charpol = -\frac{\sigma_0}{8\pi k_0}\frac{\Gamma}{\Delta_{ff'}+i\Gamma/2}$ is the complex polarizability and the irreducible tensor operator $ \hat{\tensor{\mbf{A}}}(f,f') $ is given in \erf{Eq::PolarizabilityIrrep}.

The rate of optical pumping between clock states $\ket{f,0} \rightarrow \ket{\tilde{f},0}$ is
	\begin{equation}\label{Eq::gammaff}
		\!\!\gamma_{f \!\rightarrow\! \tilde{f} }\! 
		=\!\sum_{q}\!\big| \bra{\tilde{f},\!0} \hat{W}_q^{\!\tilde{f}\!f} \!\ket{f,\!0} \big|^2\!,
	\end{equation}
where $ \hat{W}_q^{\tilde{f}f} = \sum_{f'}\frac{\Omega/2}{\Delta_{f'\tilde{f}}+i\Gamma/2}(\mathbf{e}_q^*\cdot\hat{\mathbf{D}}_{\tilde{f} f'} )(\mathbf{e}_{\rm in}\cdot \hat{\mathbf{D}}^\dagger_{f'f} ) $ are the Lindblad jump operators for optical pumping between ground levels $ f\rightarrow \tilde{f} $~\cite{deutsch_quantum_2010}. 
Each jump operator $\hat{W}_q^{\tilde{f}f}$ is associated with absorption of the probe photon polarized along $ \mathbf{e}_{\rm in} $ followed by spontaneous emission of a photon with polarization $ \mathbf{e}_q $, where $q= \{0,\pm 1\}$ labels spherical basis elements for $\pi$ and $ \sigma_\pm$ transitions.  

To find the dependence on the input field intensity, we define a characteristic photon scattering rate, $\gamma_s \equiv \frac{\Gamma\Omega^2}{4\Delta_{J_3}^2}= \frac{\sigma_0}{A_{\rm in}}\frac{\Gamma^2}{4 \Delta_{J_3}^2} \dot{N}_L $, with Rabi frequency $ \Omega=2\bra{j}|d|\ket{j'}\mathcal{E}^{(+)}_{\rm in}/\hbar $, reduced optical dipole matrix element $\bra{j}|d|\ket{j'}$, and field amplitude $ \mathcal{E}^{(+)}_{\rm in}=|\mathbf{E}_{\rm in}^{(+)}(\br')| $.
Eqs.~\eqref{Eq::gammaf} and~\eqref{Eq::gammaff} yield,
\begin{subequations}
	\begin{align}
		\!\!\!\!\!\!\!\!\!\!\!\!\!\!\!\!\!\!\!\!\!\!\!\!\!\gamma\!_{_f} &\!=\!n_g\dot{N}\!\!_L  \!\!\sum_{f'}\!\! \sigma (\Delta_{f\!f'} \!) \mathbf{u}^*_\inp(\!\br'\!\!\!_{_\perp}\!)\!\cdot\! \bra{f\!,\!0} \hat{\tensor{\mbf{A}}}(f\!,\!f') \ket{f\!,\!0} \! \cdot\! \mathbf{u}_\inp(\!\br'\!\!\!_{_\perp}\!)\\
		&\!\approx\!  \gamma_s\! \sum_{f'}\! \frac{\Delta_{J_3}^2}{\Delta_{f\!f'}^2}\!\sum_q\!\! \left| o_{jf}^{j'\!f'}C_{f'q}^{f0;1 q}\right|^2\! \mathbf{e}_q^* \!\cdot\! (\mathbf{e}_{\rm in}\mathbf{e}_{\rm in}^* )\!\cdot\! \mathbf{e}_q, 
	\end{align}
\end{subequations}
	\begin{align}
		\!\!\!\!\!\gamma\!_{_{f \!\rightarrow\! \tilde{f}}} 
		&\!\approx\! \gamma_s \!\!\sum_{f'}\!\! \frac{\Delta_{J_3}^2}{\Delta_{f\!f'}^2}\!\!\sum_q\!\! \left|\! o_{j\tilde{f}}^{j'\!\!f'}\! o_{jf}^{j'\!\!f'}\!\!C_{f'q}^{\tilde{f}0;1 q} \!C_{f'q}^{f0;1q} \!\right|^2 \!\! \mathbf{e}_q^* \!\!\cdot\!\! (\mathbf{e}_{\rm in}\!\mathbf{e}_{\rm in}^* )\!\!\cdot\! \mathbf{e}_q,
	\end{align}
where $ \sigma (\Delta_{f\!f'} ) \! =\! \sigma_0 \Gamma^2\!/\!4\Delta^2_{f\!f'}$ is the scattering cross section at the probe detuning, $ C_{f'q}^{f0;1 q}=\Braket{f'q}{f0;1q}$ are the Clebsch-Gordan coefficients, and
\begin{equation}
\big| o_{jf}^{j'f'} \big|^2=(2j'+1)(2f+2) \bigg\{
\begin{array}{ccc}
f' & 7/2 & j' \\
 j & 1 & f 
 \end{array}
 \bigg\}
\end{equation}
are the relative oscillator strengths determined by the relevant Wigner 6-$J$ symbol.

%===================APPENDIX: Equations of motion =====================%
\section{Derivation of the equations of motion for the moments} \label{Appendix::OpticalPumping}	

In this Appendix we derive the equations of motion for the correlation functions that define the metrologically relevant squeezing parameter, $\xi^2 = N_A \Delta J_3^2/\expt{\hat{J}_1}^2$.  
We seek the time evolution of the one and two-body correlation functions:
\begin{subequations}
\begin{align}
&\expt{\hat{N}_C} = \sum_n \expt{\hat{\mathbbm{1}}^{(n)}_C} \\
&\expt{\hat{J}_1} = \frac{1}{2} \sum_n \expt{\hat{\sigma}_1^{(n)}} \\
&\expt{\hat{J}_3} = \frac{1}{2} \sum_n \expt{\hat{\sigma}_3^{(n)}} \\
&\expt{\hat{J}_3^2} = \frac{\expt{\hat{N}_C}}{4} +\frac{1}{4} \sum_{m \neq n} \expt{\hat{\sigma}_3^{(m)}\otimes \hat{\sigma}_3^{(n)}}, 
\end{align}
\end{subequations}
where $\hat{\mathbbm{1}}_C \equiv \op{\uparrow}{\uparrow} + \op{\downarrow}{\downarrow}$ is the single-atom projector onto the clock states. 
To include optical pumping, we apply the following equations of motion. For a collective, single-body operator, $\hat{X} = \sum_n \hat{x}^{(n)}$, the evolution due to optical pumping is $d\expt{ \hat{X}}|_{\rm op} = \sum_n \Tr [\mathcal{D}_n[\hat{\rho}]\hat{X} ]dt = \sum_n \expt{\mathcal{D}_n\dg[\hat{x}^{(n)}]} dt$, where the map, which acts locally on atoms along the nanofiber, is given in \erf{Eq::OpticalPumpingMapSchr}.  
Two-body microscopic operators decay by optical pumping according to \cite{baragiola_three-dimensional_2014}
	\begin{align} \label{Eq::TwoBodyDecay}
		\!\!\!\!\!\frac{d}{dt}\! \expt{ \hat{x}^{(\!m\!)} \!\otimes\! \hat{y}^{(\!n\!)} } \!\Big|_{\rm op} \!\!= &\expt{ \mathcal{D}_m\dg[ \hat{x}^{(\!m\!)}] \!\otimes\! \hat{y}^{(\!n\!)} } \!+\! \expt{ \hat{x}^{(\!m\!)}\!\otimes\! \mathcal{D}_n\dg[ \hat{y}^{(\!n\!)}] },
	\end{align}
where the superscripts refer to the $m^{th}$ and $n^{th}$ atoms. 

Applying the adjoint map to the single-atom operators yields 
	\begin{subequations} \label{Eq::OperatorMap}
	\begin{align}
		\mathcal{D}\dg[\hat{\mathbbm{1}}_C] & = - \gamma_{00} \hat{\mathbbm{1}}_C +\gamma_{03} \hat{\sigma}_3 \label {Eq::Idecay} \\
		\mathcal{D}\dg[\hat{\sigma}_3] & =- \gamma_{33} \hat{\sigma}_3 +  \gamma_{30} \hat{\mathbbm{1}}_C 
\label{Eq::zdecay} \\
		\mathcal{D}\dg[\hat{\sigma}_1] & = - \gamma_{11} \hat{\sigma}_1\label{Eq::xdecay},
	\end{align}
	\end{subequations}
with rates	
	\begin{subequations} \label{Eq::DecayRates}
	\begin{align}
		\gamma_{00} 
			& = \frac{\gamma_{\uparrow}+\gamma_{\downarrow} - \gammauu-\gammaud  -\gammadd-\gammadu}{2} \label{Eq::lrate} \\
			\gamma_{03} 
			& = \frac{-\gamma_{\uparrow}+\gamma_{\downarrow} +\gammauu + \gammaud - \gammadd - \gammadu }{2}\\		
		\gamma_{33} 
			& = \frac{\gamma_{\uparrow}+\gamma_{\downarrow} - \gammauu+\gammaud  -\gammadd+\gammadu}{2}\\
			\gamma_{30} 
			& = \frac{-\gamma_{\uparrow} + \gamma_{\downarrow} + \gammauu - \gammaud - \gammadd + \gammadu }{2} \\
			\gamma_{11} 
			& = \frac{\gamma_{\uparrow}+\gamma_{\downarrow}}{2}. \label{Eq::frate}
	\end{align}
	\end{subequations}
Given Eqs. (\ref{Eq::TwoBodyDecay}, \ref{Eq::OperatorMap}), the equations for the two-body spin correlations, \erf{Eq::TwoBodySpinDecay}, follow.  Similarly, one can derive equations of motion for the remaining two-body microscopic operator correlations $ \expt{\hat{\mathbbm{1}}^{(m)}_C \otimes \hat{\mathbbm{1}}^{(n)}_C} $ and $ \expt{\hat{\mathbbm{1}}^{(m)}_C \otimes \hat{\sigma}_3^{(n)} + \hat{\sigma}_3^{(m)} \otimes \hat{\mathbbm{1}}^{(n)}_C } $ when $ m\neq n $ and from these, the macroscopic operator expectation values $ \expt{\hat{J}_3^2} $, $ \expt{\hat{N}_C^2} $, and $ \expt{\hat{N}_C\hat{J}_3} $.  As we have examined numerically, on the time scale of the QND measurement, the correlation between atom number in the clock state subspace and the pseudospin moment is weak, and one can thus treat the atom number operator in the clock state subspace as a $c$-number.
We therefore set $ \expt{\hat{N}_C\hat{J}_3}-\expt{\hat{N}_C}\expt{\hat{J}_3} = 0 $ and $ \expt{\hat{N}_C^2} - \expt{\hat{N}_C}^2 = 0 $ and define $ N_C\equiv \expt{\hat{N}_C}$. 

The equations of motion for the moments of $\jz$ are now found from the SME, \erf{Eq::SME},
	\begin{subequations} \label{Eq::J3MomentEquations}
	\begin{align} 
		\!\!d \expt{\!\jz\!} \!=& s \sqrt{\kappa} \varz \, dW - \gamma_{33} \expt{\jz}dt + \smallfrac{1}{2} \gamma_{30} N_C dt ,  \\
		\!\!d \expt{\!\jz^2\!} \!=& 2 s\sqrt{\kappa} \expt{\jz}\Delta J_3^2 \, dW \!-\! 2 \gamma_{33} \expt{\!\jz^2}dt \!+\! \smallfrac{1}{4} \big(\! 2 \gamma_{33}\!-\!\gamma_{00}\!\big) N_C dt \nn\\
		&+ \gamma_{30} \expt{\jz} N_C dt + \smallfrac{1}{2}\left(\gamma_{03} -2 \gamma_{30}\right) \expt{\jz} dt. 
	\end{align}
	\end{subequations}
The stochastic term in $d\expt{\jz^2}$ was simplified by assuming Gaussian statistics \cite{jacobs_straightforward_2006}, $\expt{\jz^3} = 3\expt{\jz^2}\expt{\jz}- 2\expt{\jz}^3$. 
Finally, the It\={o} calculus governing the stochastically evolving moments requires that differentials be taken to second order, and the evolution of the variance is given by $d \varz = d \expt{\jz^2} - 2 \expt{\jz} d \expt{\jz} - ( d \expt{\jz} )^2$. 
The equation of motion for the conditional variance, \erf{Eq::varJz}, then follows from Eqs. \eqref{Eq::J3MomentEquations}.

\end{appendix}

\end{document}